\documentclass[fleqn,usenatbib,useAMS]{mnras}

\usepackage[T1]{fontenc}
\usepackage{ae,aecompl}

\usepackage{graphicx}	
\usepackage{amsmath}	
\usepackage{amssymb}	
\usepackage{amstext}
\usepackage{amsfonts}
\usepackage{float}

\newcommand{\sncut}{15}
\newcommand{\onlyoneexp}{230}
\newcommand{\fg}{62}
\newcommand{\unknown}{11} 
\newcommand{\nocol}{73}
\newcommand{\bgcut}{18}
\newcommand{\goodstarmember}{687}
\newcommand{\goodstarmbg}{705} 
\newcommand{\starfitted}{907}
\newcommand{\ltstar}{982}
\newcommand{\lowmetmemberbgzero}{156}
\newcommand{\lowmetmemberbgmofive}{37}
\newcommand{\lowmetmemberbgmone}{8}
\newcommand{\teff}{\ensuremath{T_\textrm{eff}}}
\newcommand{\logg}{\ensuremath{\log(g)}}
\newcommand{\mh}{[M/H]}

	\title[Metallicity distribution in the Galactic Centre]{KMOS view of the Galactic Centre  -- II. 
	Metallicity distribution of late-type stars \thanks{Based on observations collected at the European Organisation for Astronomical Research in the Southern Hemisphere, Chile (60.A-9450(A)).}}

	\author[A.~Feldmeier-Krause, et al.]
	{A.~Feldmeier-Krause,$^{1}$\thanks{E-mail: afeldmei@uchicago.edu} 
   W.~Kerzendorf,$^{1}$  
    N.~Neumayer,$^{2}$ 
    R.~Sch{\"o}del,$^{3}$ 
    \newauthor
    F.~Nogueras-Lara,$^{3}$  
	T.~Do,$^{4}$     
    P.~T.~de~Zeeuw,$^{1,}$$^{5}$ 
    and 
    H.~Kuntschner$^{1}$ 
\\
		$^{1}$European Southern Observatory (ESO), Karl-Schwarzschild-Stra{\ss}e 2, D-85748 Garching, Germany\\
$^{2}$Max-Planck-Institut f{\"u}r Astronomie, K{\"o}nigsstuhl 17, D-69117 Heidelberg, Germany\\  		     
$^{3}$Instituto de Astrof\'{i}sica de Andaluc\'{i}a (CSIC), Glorieta de la Astronom\'{i}a s/n, E-18008 Granada, Spain   \\  
$^{4}$UCLA Galactic Center Group, Physics and Astronomy Department, UCLA, Los Angeles, CA 90095-1547, USA \\ 
$^{5}$Sterrewacht Leiden, Leiden University, Postbus 9513, NL-2300 RA Leiden, The Netherlands 
}

\date{Accepted 2016 September 12. Received 2016 September 9; in original form 2016 May 20}

\pubyear{2016}

\begin{document}
\label{firstpage}
\pagerange{\pageref{firstpage}--\pageref{lastpage}}
\maketitle

\begin{abstract}
Knowing the metallicity distribution of stars in the Galactic Centre has important implications for the formation history of the Milky Way nuclear star cluster. However, this distribution is not well known, and is currently based on a small sample of fewer than 100 stars. 
We obtained  near-infrared $K$ band spectra of more than 700 late-type stars in the central 4\,pc$^2$ of the Milky Way nuclear star cluster with the integral-field spectrograph KMOS (VLT).  We analyse the medium-resolution spectra using a  full-spectral fitting method employing the G{\"o}ttingen Spectral library of synthetic PHOENIX spectra.  The derived stellar metallicities range from metal-rich $\mh$\textgreater$+$0.3\,dex to  metal-poor  $\mh$\textless$-$1.0\,dex, with a fraction of 5.2$^{+6.0}_{-3.1}$ per cent metal-poor  ($\mh$\,$\leq$\,$-$0.5\,dex) stars.
 The metal-poor stars are distributed over the entire observed field. The origin of  metal-poor stars remains unclear.  They could originate   from infalling globular clusters. 
For the  metal-rich stellar population  ($\mh$\,\textgreater\,0\,dex) a globular cluster origin can be ruled out. 
As there is only a very low fraction of metal-poor stars in the central 4\,pc$^2$ of the Galactic Centre, we believe that our data can discard a scenario in which the Milky Way nuclear star cluster is purely formed from infalling globular clusters.
\end{abstract}




\section{Introduction}
A dense concentration of stars is located within the central  10~pc of our Galaxy. These stars form the Milky Way's nuclear star cluster.  The centre of this star cluster hosts the Milky Way's supermassive black hole, Sgr~A*. It is unclear how the nuclear star cluster and the supermassive black hole formed and evolved.  Many other galaxies host either a  nuclear star cluster,  a central supermassive  black hole, or both \citep[e.g.][]{boker02,cote06,anil08,graham09,nadine_ncbh12}. 

Two main formation scenarios for nuclear star clusters have been proposed. The stars may have formed `\textit{in situ}', i.e., in the  centre of the galaxy where they are located now \citep[e.g.][]{loosesf82,2008ApJ...684L..21S,anilon4244_08,pflamm09}. 
The second possibility is that the stars formed in clusters further out. Star clusters might have migrated to the centre of the galaxy to form the nuclear star cluster \citep[e.g.][]{tremaine75,capuzzo08,2011ApJ...729...35A,gnedin13,antonini13}. In particular, massive globular clusters may have fallen into galactic nuclei through dynamical friction and thus have seeded nuclear star clusters \citep[e.g.][]{antonini12}. Studying the stellar populations of the nuclear star cluster will shed light on the formation scenario, possibly distinguishing between the presented scenarios. In particular, the stellar metallicity distribution is useful to infer the formation history of the nuclear star cluster. 
A narrow distribution of age and metallicity would suggest that all stars formed from the same gas cloud in one single burst of star formation. On the other hand, a  spread in age and metallicity would reveal several star formation events, or even continuous star formation.  An inhomogeneous distribution of the metallicity over the nuclear star cluster might be produced by the migration of star clusters with different intrinsic metallicities  from the outskirts to the centre of the galaxy \citep{perets14}. In particular, if globular clusters have contributed significant mass to a nuclear star cluster, then we would expect to find a considerable fraction of low-metallicity stars  in it, as globular cluster stars have typically low metallicities, [$Fe/H$]. In the Milky Way globular clusters, [$Fe/H$]  ranges from $-2.37$  to 0.0\,dex with a median at [$Fe/H$]=$-$1.32\,dex   \citep[2010 edition]{harris}. Only about 10\,per cent of the Milky Way globular clusters have metallicities [$Fe/H$]  greater than --0.5\,dex \citep{bica06}. 

Due to its proximity, the Milky Way nuclear star cluster is the perfect laboratory to study the composition  and structure of a nuclear star cluster. The cluster is at a distance of approximately 8\,kpc \citep{ghez08,gillessen09s2,chatzopoulos}. 
It has a mass of approximately 2.5$\times$10$^7$\,M$_\odot$ \citep{sb,isaacanja} and a half-light radius of 4.2~pc \citep{sb}. 
Many of the stars observed in the Galactic Centre are cool late-type stars. They are distributed throughout the cluster \citep{kmoset}. 
Most of the cool stars are red giants that formed several Gyr ago, though some late-type stars are  younger supergiant stars \citep{blum03,pfuhl11}. Several red giants of intermediate age (50--500\,Myr) have been found as well \citep{2016A&A...588A..49N}. With the exception of short-lived O/B stars and bright A-stars, main-sequence stars are too faint to be detected and studied with current instrumentation in the crowded Milky Way centre.

Despite being the arbiter for a number of important questions, there exist only few measurements of  iron abundance and metallicity  in the Galactic Centre. Lately, the \textit{Gaia}-ESO Survey and the Apache Point Observatory Galactic Evolution Experiment (APOGEE) measured metallicity and abundances for several thousand stars of the Milky Way \citep{2014A&A...572A..33M,2015AJ....150..148H}. However, APOGEE observed only few stars in the Galactic Centre.
Measurements of the  \textit{Gaia}-ESO Survey at visual wavelengths are hindered by   high  extinction  \citep[$A_V$\,=\,30\,mag;][]{scoville03} towards the centre of the Milky Way.  Infrared spectroscopy is required, preferentially in the near-infrared $K$ band, where extinction decreases to about $A_K$\,=\,2.5\,mag \citep[e.g.][]{2001A&A...376..124C,rainer10}. 
\cite{ramirez00} and \cite{rydechem14}  measured the iron abundance of fewer than 20 stars in  the Galactic Centre using high-resolution spectroscopy and found a mean iron abundance near solar, $\langle$[$Fe/H$]$\rangle$\,=\,+0.1\,dex. 
 Recently,  \cite{dolowfe} measured the overall metallicity $\mh$\space on a larger sample of 83 stars and found a large spread, from $\mh$\,$\la$\,$-1.0$\,dex to $\mh$\,$\ga$\,+0.5\,dex.  Their sample is concentrated in the central 1\,pc,  where adaptive optics is most useful. 
 Further out, in the inner Galactic bulge, \cite{schultheis15} found also evidence for metal-poor K/M giants with $\mh$\,=\,$-$1\,dex. 
The $\alpha$-elements of these metal-poor stars  seem to be enhanced \citep{schultheis15}.  Concerning metal-rich stars ([$Fe/H$\textgreater0\,dex)  in the Galactic Centre,  \cite{cunha07} and \cite{rydechem14} found enhanced Ca abundances  (0\,dex$\la$[$Ca/Fe$]$\la$+0.5\,dex), while \cite{rydechem14} and \cite{2016AJ....151....1R} found lower Mg and Si abundances  ($-$0.2$\la$$\left[ \alpha/Fe \right]$$\la$+0.2\,dex). 

A systematic metallicity study  of a large sample of stars  beyond the central $r$\,=\,0.5\,pc of the Milky Way nuclear star cluster has been missing so far.
 In this study, we present the spectra of more than 700 late-type stars located in an area \textgreater4~pc$^2$ of the Galactic Centre. The spectra were obtained with the integral-field spectrograph KMOS \citep{kmos} in the near-infrared $K$ band. We measure the stellar parameters effective temperature $\teff$, and overall metallicity $\mh$ with full-spectral fitting for more than 700 stars and study their spatial distribution.
The outline of this paper is as follows: we present the data set in Section \ref{sec:sec2}. We outline the full-spectral fitting routine and the error estimation in Section \ref{sec:sec3}. In Section \ref{sec:sec4}, we present our results and discuss them in Section \ref{sec:sec5}. The conclusion follows in Section~\ref{sec:sec6}.

\section{Data set}
\label{sec:sec2}
\subsection{Observations and data reduction}
Our spectroscopic observations were performed with KMOS  on 2013 September 23,  during the KMOS science verification at VLT-UT1 (Antu). We observed an area of  2\,700\,arcsec$^2$, which corresponds to approximately 4\,pc$^2$ at a distance of 8\,kpc. The field extends over  64.9\,arcsec $\times$ 43.3\,arcsec, centred on $\alpha$\,=\,266\fdg4166 and $\delta$\,=\,$-$29\fdg0082, with a gap of 10.8\,arcsec $\times$ 10.8\,arcsec in the Galactic North-East direction, since one of the 24 KMOS integral-field units (IFUs) was inactive  \citep[IFU 13, see fig. 1 of][]{kmoset}.  

The spectra were taken in the $K$ band   ($\sim$19\,340 -- 24\,600\,$\AA$). 
The KMOS scale is $\sim$2.8\,$\AA$\,pixel$^{-1}$ in the spectral direction. The spatial  scale is 0.2\,arcsec\,pixel$^{-1}$\,$\times$\,0.2\,arcsec\,pixel$^{-1}$. 
We observed the field twice with 100\,s exposure time each. 
For sky subtraction, we made an offset to a dark cloud (G359.94+0.17, $\alpha$\,$\approx$\,266\fdg2, $\delta$\,$\approx$\,$-$28\fdg9;   \citealt{dutra_darkcloud_01}).  Further, we observed B dwarfs  for telluric corrections. 
Data reduction was performed with the ESO pipeline and  standard recipes for dark correction, flat-fielding,  wavelength calibration, and illumination correction. For sky subtraction we used the method by \cite{ohsky}, which is implemented in the pipeline. We removed cosmic rays with the method by \cite{lacosmic}. 
For further details on the data reduction, we refer to \cite{kmoset}.

We extracted spectra of more than 1\,000 individual stars with \textsc{PampelMuse} \citep{pampel}. The spectra have a  formal signal to noise  \textgreater10.   In \textsc{PampelMuse}, stars are deblended with a point spread function (PSF) fitting technique. Spectra from stars are also extracted when the stars' PSFs are centred outside of the field of view of the IFU. Therefore, we have more than two exposures for some of the stars. 

\setcounter{table}{0}
\begin{table*}
 \centering
  \begin{minipage}{115mm}
\caption{Spectral libraries in the $K$ band}
 \label{tab:splib}
\begin{tabular}{@{}lccccc@{}}
\noalign{\smallskip}
\hline
\noalign{\smallskip}
Library& Resolution&Spectral &Luminosity& Spectral region&Number\\
&	($R=\lambda/\Delta\lambda$)&	type&class	&($\AA$)&of stars\\
 \noalign{\smallskip}
\hline
\noalign{\smallskip}
 \cite{wallace}			&$\geq$45\,000	&F8--M8		& I--V	& 20\,195--23\,960	&9 \\
  \cite{wallacemr}		&3\,000		&O4--M7	&I--V	&20\,202--24\,096	&61\\
\cite{winge}, NIFS v1.5	&6\,000		&G8--M3	& I--III	&20\,700--24\,700	&10	\\
\cite{winge}, NIFS v2.0	&6\,000		&G8--M3	& I--III	& 20\,200--24\,300	&13	\\
 \cite{winge}, GNIRS		&18\,000		&F7--M0	& II--V	&21\,800--24\,250	&16	\\
 \hline 
\end{tabular}
\end{minipage}
\end{table*}

\subsection{Spectral resolution}
\label{sec:lsf}
The spectral resolution of KMOS varies spatially for the 24 different IFUs \citep[e.g.][]{2015ApJ...805..182G}. We  measured the line-spread function of each IFU separately on the  reconstructed sky data cubes in order to create a resolution map for each IFU. 
The line-spread function is reasonably well described with a Gaussian, and we  fitted  
three different sky lines in the wavelength region of $\lambda$\,=\,21\,900--22\,400\,$\AA$ for all 23 used IFUs separately.  
We calculated  average resolution maps from four sky exposures for each IFU, and smoothed the maps using a Gaussian with width $\sigma$ = 1\,pix, roughly corresponding to the seeing during the observations. We found that the spectral resolution $R$ = $\lambda / \Delta \lambda$ varies between 3\,310 and 4\,660 for the 23 different active  IFUs on the three different KMOS detectors. On a single IFU, the spectral resolution has a standard deviation of about 30--150. The standard deviation of the spectral resolution $R$ over all IFUs is 300, this is 7 per cent of the mean value. We used the respective $R$ values for fitting the stellar spectra. 

\subsection{Data selection}
For this paper we only regard late-type stars, i.e., cool stars with molecular CO absorption lines at $\lambda$\,$\geq$\,22\,935\,$\AA$. Spectra of  early-type stars are presented in \cite{kmoset}. We  visually inspected the spectra and identified 982 stars as late-type stars based on the prominent CO lines.

In addition to the spectroscopy, we have photometry obtained with HAWKI-I and NACO from \cite{rainer10} and Nogueras-Lara et al. (in preparation) in the $J$ (HAWK-I), $H$ (HAWKI-I and NACO), and $K_S$ (HAWKI-I and NACO)  bands. Since the brightest stars can be saturated in the HAWKI-I and NACO images, we complemented the photometry with the SIRIUS catalogue \citep{shogo06} for six late-type stars. For two further bright late-type stars with neither  HAWK-I, NACO nor SIRIUS photometry due to saturation, we used Two Micron All Sky Survey   \citep[2MASS;][]{2mass} photometry.

We corrected the photometry for dust extinction using the extinction map and extinction law of \cite{rainer10}, which was derived with NACO data.  This extinction map  covers about  70 per cent of our stars. For stars outside the field of view of the  \cite{rainer10} extinction map, we used the extinction map of Nogueras-Lara et al. (in preparation) derived from  HAWK-I data. It  has lower spatial resolution, but covers the entire field of view of the KMOS data. 

The $H-K_S$ colour can be used to identify foreground stars. The intrinsic $H-K_S$ colour ranges from about $-$0.13  to +0.38\,mag \citep{do13,rainer_review14}. 
We assumed that stars with significantly bluer extinction-corrected $(H-K_S)_0$ colour are overcorrected foreground stars. We classified a star as foreground star when $(H-K_S)_0$\,\textless\,$-$0.5\,mag. 
Including zero-point uncertainties, the uncertainty of the $H$- and $K_S$-band magnitudes in our photometry is less than 10 per cent, so $(H-K_S)_0$ \textless $-$0.5\,mag corresponds to a 3$\sigma$ exclusion criterion.
We excluded  \fg\space stars with this criterion. For \unknown\space stars, we do not have photometry in both bands. We excluded these \unknown\space stars  in our later analysis, since they might be foreground stars. In our selected data set, the extinction corrected $K_S$-band magnitudes range  from 4.91 to 13.56\,mag, with a median of 10.49\,mag. For a colour--magnitude diagram of the data set, we refer to \cite{kmoset}.

\subsection{Spectral indices}
\label{sec:tco}

Spectral indices in the $K$ band are correlated with the spectral type of a star, and can be used for a rough spectral classification and effective temperature estimate \citep[e.g.][]{ivanov,silva}. The most prominent spectral features of cool stars are the CO absorption lines ($\lambda$\,$\geq$\,22\,935~$\AA$) and the \ion{Na}{I} doublet at 22\,062 and 22\,090~$\AA$.  
In this section, we calibrate  the CO equivalent width ($EW_\mathrm{CO}$) and the Na equivalent width  ($EW_\mathrm{Na}$) on a spectral library and present the  measurements on our data set. The $EW_\mathrm{CO}$  and $EW_\mathrm{Na}$ measurements confirm our classification as late-type stars quantitatively. Further, we obtain some constraints on the effective temperature $\teff$ and surface gravity $\logg$ that are useful as priors for the full-spectral fitting. 

\subsubsection{Calibrating spectral indices with  a spectral library}

We measured the spectral indices $EW_\mathrm{CO}$ and $EW_\mathrm{Na}$ using different spectral libraries with known spectral types for calibration. The  spectral libraries  are listed in Table \ref{tab:splib}. We used the spectral index definitions of \cite{frogel}. 
All spectra were degraded to the same spectral resolution of the KMOS data before we computed the indices. The spectral resolution of the KMOS detector varies for the different IFUs (see Section~\ref{sec:lsf}). We tested the effect of the spectral resolution by degrading the spectral library to 
 $R$\,=\,3\,000, and found no systematic difference in the result compared to  the spectral indices obtained with $R$\,=\,4\,350. The  difference is less than 1 per cent for all indices. 
This test shows that equivalent width measurements are robust under moderate spectral resolution variations. Since the results with  $R$\,=\,3\,000 are consistent with the results at KMOS resolution, we  included   spectra from \citet[$R$\,=\,3\,000]{wallacemr}\footnote{We excluded stars that are also in the sample of \cite{wallace}, and the star HR8530/HD212320, which was classified as M6III star by \cite{wallacemr}, but was listed as  G6III star by \cite{mcwilliam90}.} to obtain a larger sample. 
We also tested the influence of the continuum shape on  index measurements by reddening the library spectra with the mean extinction in our field of view, $A_{K_S}$\,=\,2.7\,mag. For the CO and Na equivalent width measurements  the effect is less than 1 per cent, though it is about 7.3 per cent when we instead compute  the $D_\mathrm{CO}$ index as defined by \cite{marmol}. 

Giant stars with luminosity class II--IV have a mean value for $EW_\mathrm{CO}$ of approximately 13\,$\AA$, and the maximum value is about 25\,$\AA$. We expect that the majority  of the stars in our data set are giants. 
However, we also computed  $EW_\mathrm{CO}$ for the supergiant and dwarf stars in our spectral library. Supergiants have larger values of $EW_\mathrm{CO}$, with a mean at  20\,$\AA$, and values up to 33\,$\AA$. Since no giant star has a larger value of $EW_\mathrm{CO}$ than 25\,$\AA$, we can assume that  stars in our data set with $EW_\mathrm{CO}$ \textgreater 25\,$\AA$ are either  M-type supergiants, or have a higher metallicity, as CO absorption strength also increases  with metallicity (e.g. \citealp{1992AJ....103..163H} for [$Fe/H$]\,\textless\,0\,dex).    The mean value of $EW_\mathrm{Na}$ is about 2\,$\AA$ for giant stars and about 3\,$\AA$ for supergiants.  We further found that all stars with a detectable  CO line, from dwarfs over giants to supergiants, have effective temperatures $\teff$\,$\la$\,6\,000\,K. Since all the stars in our data set have CO lines, their temperatures are probably $\la$ 6\,000\,K.
We derived a new $\teff$ -- $EW_\mathrm{CO}$ relation for giant stars. The results are presented in Appendix~\ref{sec:tcorel}. 

\subsubsection{Late-type star classification }
\label{sec:ltclass}
To verify our classification as late-type stars, we measured spectral indices on our data set. We  computed the  equivalent widths of the first CO band head  
and the  \ion{Na}{I} doublet 
as defined by \cite{frogel}.
Before computing the spectral indices, we measured the radial velocities  of the stellar spectra and shifted the spectra to  rest wavelength. To obtain the velocities, we used  the \textsc{IDL} routine \textit{pPXF} \citep{ppxf} as in \cite{isaacanja}, with the high resolution spectra of  \cite{wallace} as templates. The wavelength range of the fit was from 21\,500 to 23\,900\,$\AA$.  

The late-type stars have mean values of $EW_\mathrm{CO,LT}$\,=\,18.30\,$\AA$ and $EW_\mathrm{Na,LT}$\,=\,4.60\,$\AA$. The stars we classified as early-type O/B stars have smaller equivalent widths, with a mean value of $EW_\mathrm{CO,O/B}$\,=\,$-0.76$\,$\AA$   and $EW_\mathrm{Na,O/B}$\,=$\,0.47$\,$\AA$ \citep[see also][]{kmoset}. There is a clear distinction between the distribution of the $EW_\mathrm{CO}$ of the early- and late-type stars, which confirms our visual classification.
Our data set contains 66 stars with $EW_\mathrm{CO}$\,\textgreater\,25\,$\AA$, which  might suggest that these stars are supergiants. Among those stars, 47 are rather bright   ($K_{S,0}$\,$\leq$\,10\,mag) giving further support to those objects being supergiants at the Galactic Centre.
The CO strength in Galactic Centre supergiants is normal in comparison  to disc stars  \citep{blum96}. 

The \ion{Na}{I} doublet lines are rather strong for most stars in the data set, with a mean value of $EW_\mathrm{Na}$\,=\,4.60\,$\AA$, while the giant  stars in the spectral library have $EW_\mathrm{Na}$\,$\leq$\,4.5\,$\AA$.  \cite{blum96} showed that stars in the Galactic Centre have higher \ion{Na}{I} and \ion{Ca}{I}  line strengths compared to disc stars with similar CO strengths. Also the high-resolution spectra studied by \cite{cunha07} and \cite{rydechem14} have high [$Ca/Fe$] abundances. We show three example spectra in Fig. \ref{fig:fourspec}. The \ion{Na}{I} and \ion{Ca}{I} line regions are marked as grey shaded areas.  
Na is produced in massive stars  in SN~II \citep{2006ApJ...653.1145K} and in intermediate-mass AGB stars \citep{2010MNRAS.403.1413K}. 
 In contrast to Na, Ca is an $\alpha$-element. The  \ion{Na}{I}  lines are blended with other elements, e.g. Sc, Si, Fe, V, and CN  lines \citep{wallace}. One or several of these elements might be enhanced and produce the large \ion{Na}{I} equivalent width.  

\section{Full-spectral fitting}
\label{sec:sec3}
We fitted the spectra of our KMOS data set using the \textsc{StarKit} code \citep{starkit} also used by \cite{dolowfe}. This code interpolates on a grid of synthetic spectra and utilizes the Bayesian sampler \textsc{MultiNest} \citep{multinest,multinestpy}. In the following, we give a short outline of our assumptions and used parameters. For details on the Bayesian sampling procedure we refer to \cite{dolowfe}. 

\subsection{Fitting method and assumptions}
We fitted the effective temperature  $\teff$, metallicity  $\mh$, surface gravity $\logg$, and radial velocity $v_z$.  [$\alpha/Fe$] was fixed to zero and stellar rotation was ignored. Most of the stars in our data set are red giants, which have too low rotational velocities to be resolved with our data  \citep[\textless\,10\,km\,s$^{-1}$;][]{oke54,gray89}.
Our spectral grid consists of  the model spectra from the PHOENIX spectral library \citep{husserphoenix}, with stellar parameters in the range $\teff$\,=\,[2\,300 K; 12\,000 K] with a step size of $\Delta \teff$\,=\,100\,K, $\mh$\,=\,[$-$1.5\,dex, +1.0\,dex], $\Delta \mh$\,=\,0.5\,dex, $\logg$\,=\,[0.0\,dex, 6.0\,dex], $\Delta \logg$\,=\,0.5\,dex.  The synthetic spectra have a resolution of $R$\,=\,500\,000. We convolved the spectra to the same spectral resolution as the KMOS spectra. We took the different spectral resolutions obtained on the 23 different IFUs into account  (see Section \ref{sec:lsf}).
The prior of the metallicity  $\mh$ was set uniform in the range [$-$1.5\,dex, +1.0\,dex].  The surface gravity can be constrained by the $K_S$-band magnitude, as shown by \cite{dolowfe}. Stars brighter than $K_S$\,=\,12\,mag have a lower surface gravity. Therefore, we set the uniform priors for $\logg$ in the range 2.0\,dex\,\textless\,$\logg$\,\textless\,4.5\,dex for $K_{S,0}$\,$\geq$\,12~mag and to 0.0\,dex\,\textless\,$\logg$\,\textless\,4.0\,dex for $K_{S,0}$\,\textless\,12~mag.  
In addition, we chose 0.0\,dex\,\textless\,$\logg$\,\textless\,2.0\,dex for all stars with $K_{S,0}$\,$\leq$\,10\,mag and $EW_\mathrm{CO}$\,\textgreater\,25\,$\AA$, and   0.0\,dex\,\textless\,$\logg$ \textless\,4.0\,dex for all stars with $K_{S,0}$\,\textgreater\,10\,mag and $EW_\mathrm{CO}$\,\textgreater\,25\,$\AA$, as these stars are potentially supergiants (see Section~\ref{sec:tco}).

The continua of the spectra were not normalized or straightened. Still, we did not use the continuum shape to constrain the effective temperature, as  interstellar dust and extinction also affect the  continuum shape and might  bias our  results for the stellar parameters. We fitted the spectral continuum with a fifth degree polynomial function, to minimize the difference between the observed  and model spectrum. The model spectra were multiplied with the polynomial to match the continuum shape of the data. 

\begin{figure*}
    \includegraphics[width=0.95\textwidth]{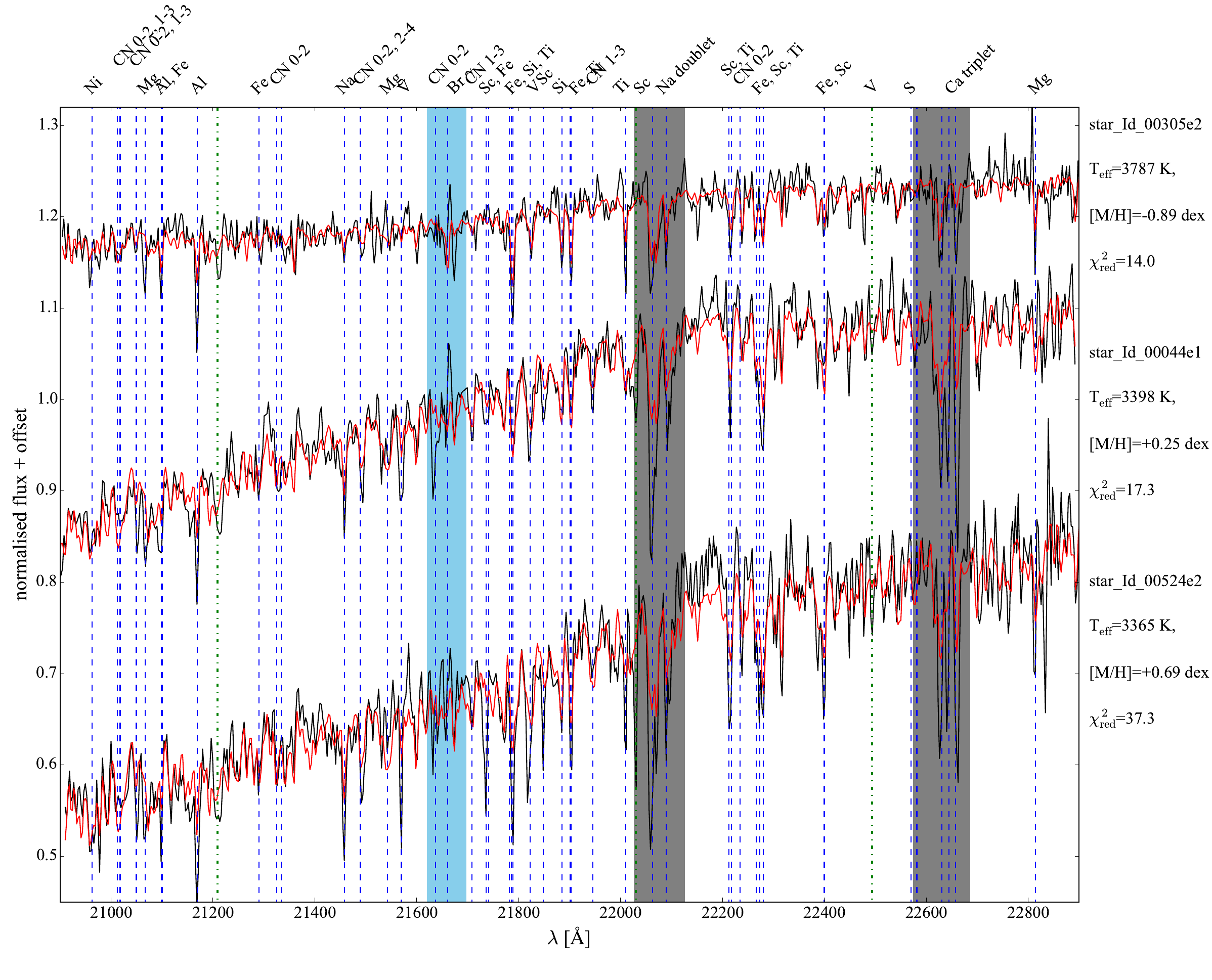}
  \caption{Spectra (black) of three different stars with metallicities $\mh$\,=\,$-$0.89,  +0.25, and +0.69\,dex. The red lines denote the best-fitting spectrum. The fits obtained  similar values for $\logg$, between 0 and 0.34\,dex, and $\teff$, between 3\,360 and 3\,790\,K. The spectra are shifted to rest wavelength. Blue dashed lines denote known spectral lines, labelled on top of the plot, green dot--dashed lines  spectral lines which appear in the observed spectra, but are not seen in the model spectra. Grey shaded areas were excluded from the fit, the light blue shaded area was excluded for spectra with strong contamination of the Brackett-$\gamma$ gas emission line.  }
 \label{fig:fourspec}
\end{figure*}

We  fitted the spectra in the wavelength range $\lambda$\,=\,[20\,900\,$\AA$, 22\,900\,$\AA$], i.e., we excluded the molecular CO absorption lines from the stellar parameter fit. \cite{dolowfe} showed that fitting the spectrum in the CO line region can introduce significant biases in $\logg$ and $\mh$.  The reason is probably that molecular lines are not as reliable as atomic lines in synthetic spectra. Since the Na and Ca lines of Galactic Centre stars are strong compared to normal disc stars (see Section \ref{sec:ltclass} and \citealt{blum96}), we excluded these lines from the stellar parameter fit. In particular, we excluded  the wavelength regions  $\lambda$\,=\,[22\,027\,$\AA$, 22\,125\,$\AA$] and $\lambda$\,=\,[22\,575\,$\AA$, 22\,685\,$\AA$]. However, we included the \ion{Na}{I} and \ion{Ca}{I} lines in the fit of the radial velocity $v_z$. 
For spectra with strong Brackett-$\gamma$ gas contamination emitted by the minispiral \citep[for a Brackett-$\gamma$ flux map extracted from our data see][]{kmoset}, we also excluded the region $\lambda$\,=\,[21\,621\,$\AA$, 21\,696\,$\AA$] from the fit. Fig. \ref{fig:fourspec} shows three spectra in the fitted wavelength region.  The excluded \ion{Na}{I} and \ion{Ca}{I} line regions are shaded in grey. The Bracket-$\gamma$ line region, which was excluded for spectra with strong gas contamination, is shaded in light blue.  Even though we excluded these regions, there are several other absorption lines in the spectra, some of them are marked with blue vertical dashed lines and labelled on top of the plot. The list of absorption lines  we labelled is not complete. There are  more lines visible in the spectra, and these lines affect the result of the fit.


\subsection{Error estimation}

\begin{figure*}
    \includegraphics[width=0.95\textwidth]{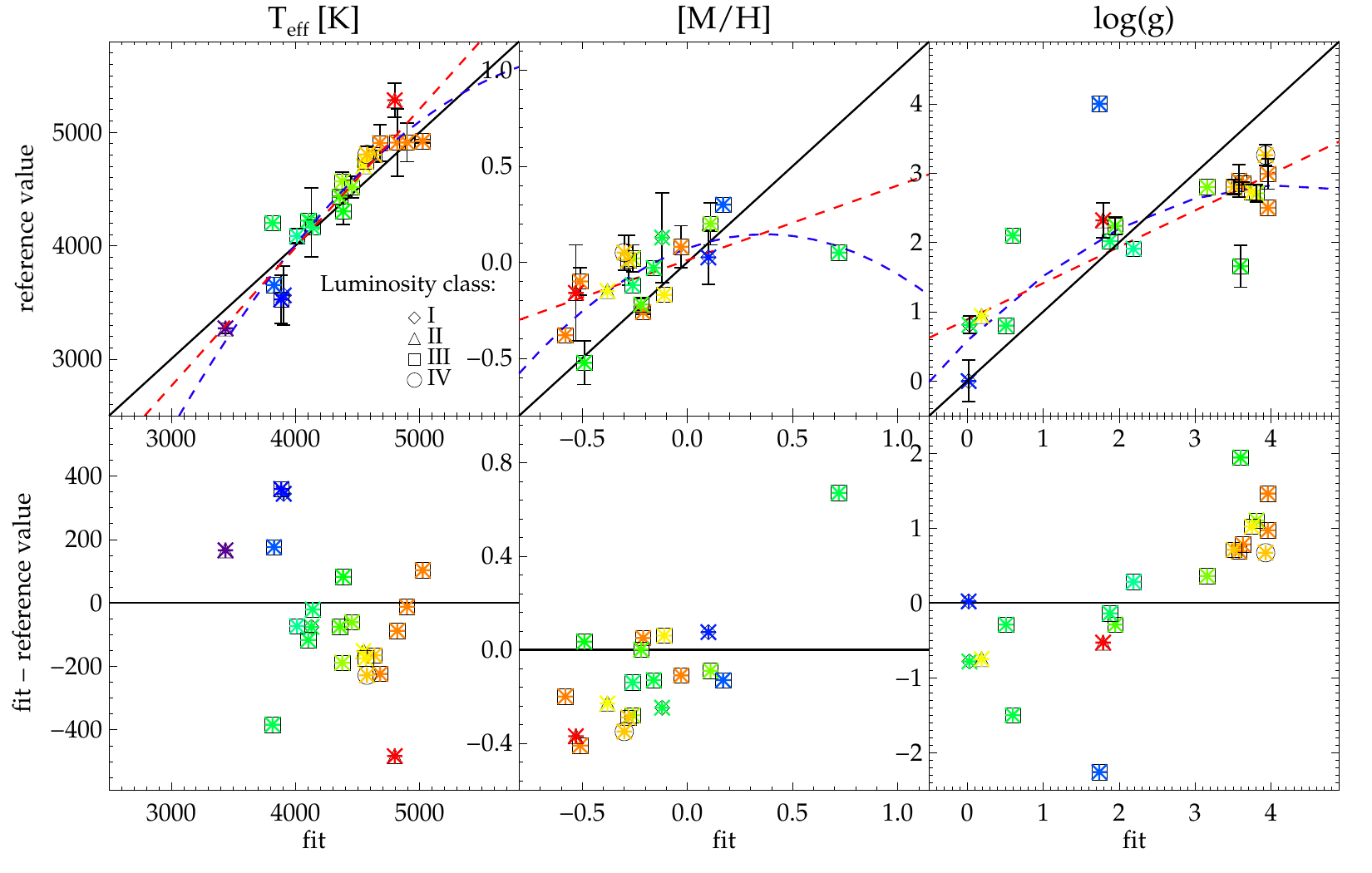}
  \caption{Comparison of the best-fitting stellar parameters with the reference values of stars from the spectral library. Left-hand panel: $\teff$ in K for 22 stars; middle panel: $\mh$\space in dex for 19 stars; right-hand panel: $\logg$ in dex  for 20 stars. Upper panel: fit result plotted against the reference value, lower panel: fit result plotted against residual  (fit result -- reference value). The different symbols denote different luminosity classes, colours mark the same star. The black line is the one-to-one line, the red dashed line is a linear fit to the data, the blue dashed line  a quadratic  fit. }	
 \label{fig:refcomp}
\end{figure*}
\subsubsection{Statistical uncertainties}
\label{sec:sigmult}
We fitted the spectra of  \starfitted\space stars. We selected spectra with high fitted signal to noise [S/N$_\text{fit}$=median(spectrum)/$\sigma$(residual spectrum)\textgreater\sncut, where the residual spectrum equals observed spectrum minus best-fitting spectrum], and with low 1$\sigma$ uncertainties ($\sigma_{ \teff}$\textless250\,K, $\sigma_{\mh}$\textless0.25\,dex, $\sigma_{\logg}$\textless 1\,dex). 
For stars with several exposures, we took the mean of the stellar parameters obtained by the  individual fits as our measurement. In many cases, the standard deviation of the results from fitting the single exposures of the same  star is larger than the 1$\sigma$ fit uncertainties. For these stars, we used the standard deviations as statistical  uncertainties. 
We estimated the uncertainties of \onlyoneexp\:stars with only one exposure from the median uncertainties of the other fits. 
The mean statistical uncertainties, either directly from the fits or the standard deviations, are  $\sigma_{\teff}$\,=\,83\,K, $\sigma_{\mh}$\,=\,0.11\,dex, $\sigma_{\logg}$\,=\,0.16\,dex, and $\sigma_{v_z}$\,=\,2.9\,km~s$^{-1}$.

One of the reasons for different results of single exposures of the same star are so-called fringes or ripples \citep{reflextut}. Fringes are visible as wavy pattern in the continuum of the spectrum. Fringes can arise  when the star is located at the edge of the IFU, or is spatially  undersampled. Fortunately, stars that are close to the edge of the IFU have often more than two exposures. The standard deviation of the individual measurements gives a good estimation  for the uncertainty of the parameter fit. 
 In some cases, the standard deviation is very high.  We conclude that the fit did not work properly for at least one of the exposures in such a case. 
We excluded stars with high statistical uncertainties. 

\begin{figure*}
    \includegraphics[width=0.95\textwidth]{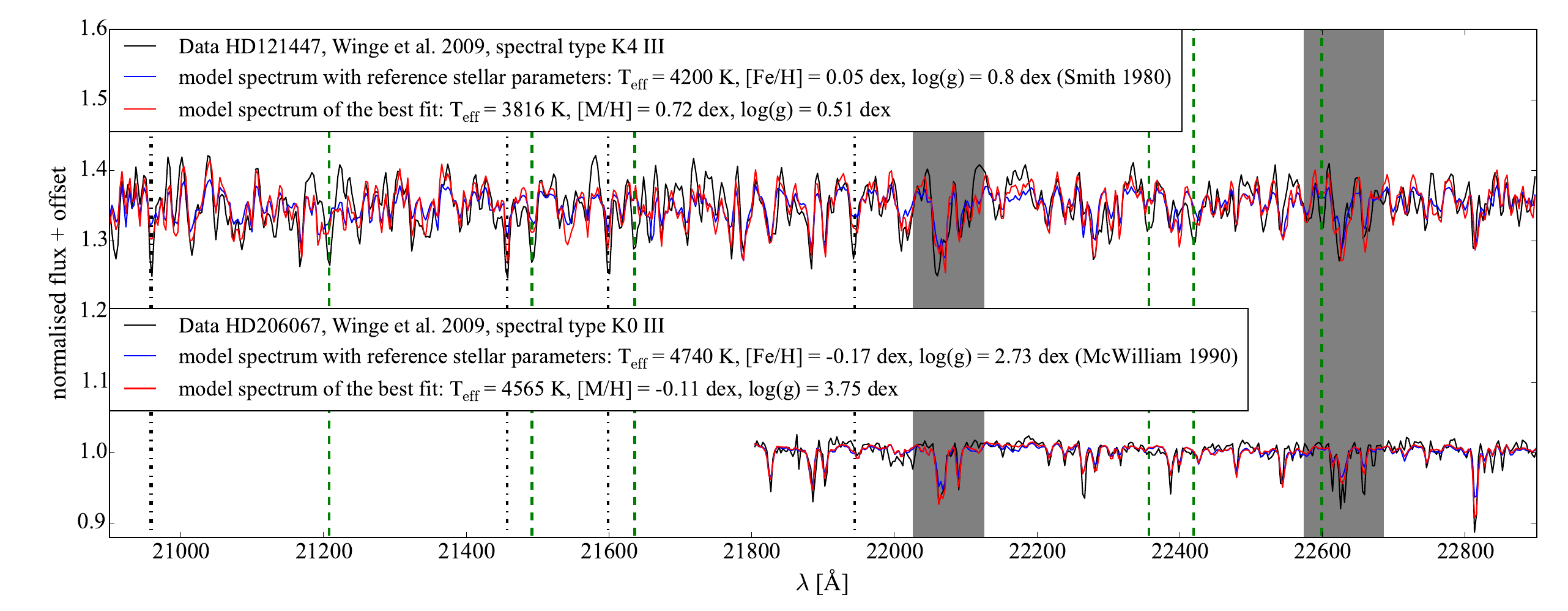}
  \caption{Top: NIFS spectrum of HD121447 \citep[black]{winge}, a PHOENIX \citep{husserphoenix}  model spectrum with the reference values found by \citet[blue]{smith84}, and a model spectrum with the best-fitting results we obtained (red, S/N$_\text{fit}$=44). Vertical dashed green lines denote spectral lines that are weaker  in the model spectra, black dot--dashed lines denote spectral lines that are better recovered by a spectrum with  higher metallicity.   Bottom: GNIRS spectrum of HD206067 \citep[black]{winge}, our results (red, S/N$_\text{fit}$=112) are in good agreement with the   reference values found by \citet[blue]{mcwilliam90}.  Grey contours mark regions that were excluded from the fit.}
 \label{fig:refspec}
\end{figure*}

\subsubsection{Systematic uncertainty of  synthetic model spectra}
\label{sec:fitlib}
We determined systematic uncertainties by applying the full-spectral fitting routine on stars from a spectral library. We selected stars from \cite{wallace} and \cite{winge} for which  previous measurements of the stellar parameters are listed in \cite{pastel}. All the spectra were convolved to the KMOS spectral resolution. The GNIRS stars from \cite{winge} cover a smaller wavelength range than our fitting interval, so we fitted those stars from $\lambda$\,=\,[21\,800$\,\AA$; 22\,900$\,\AA$]. 
We put  constraints on $\logg$ based on the luminosity class \citep[see also][]{dolowfe}. Giant stars with luminosity class III were constrained to the interval $\logg$\,=\,[0.5\,dex, 4\,dex], supergiants and bright giants (luminosity class I and II) had $\logg$\,=\,[0\,dex, 2\,dex]. Subgiants (luminosity class IV) were constrained to $\logg$\,=\,[2\,dex, 5.5\,dex]. We did not include main-sequence stars (luminosity class V), since we do not expect any dwarf stars in our  data  set except  foreground stars.  Even the faintest star in our data set with $K_S$\,=\,16.9\,mag is too bright to be a cool main-sequence star in the Galactic Centre, whereas hot   main-sequence stars (O/B/A-type) are not late-type stars and therefore excluded from the data set.
In total, we fit 22 reference spectra with known $\teff$, for 19  reference spectra we also know $\mh$, and for 20 stars $\logg$. 
We fitted each spectrum twice,  we computed the standard deviation of the residual spectrum  after the first fit and used it as uniform noise spectrum in the second fit. 
We show a comparison of the fitted results to the reference results in Fig.~\ref{fig:refcomp}. The values of $\teff$ are in good agreement, the mean and standard deviation of the fit residuals are $\langle \Delta\,\teff \rangle$\,=\,$-$58\,K, $\sigma_{\Delta \teff}$\,=\,205\,K,   $\langle \Delta\,{\mh}  \rangle$\,=\,$-$0.1\,dex, $\sigma_{\Delta \mh}$\,=\,0.24\,dex, $\langle \Delta\,{\logg} \rangle$\,=\,0.2\,dex, $\sigma_{\Delta \logg}$\,=\,1.0\,dex. To some extent, the scatter and offsets are caused by the different methods and assumptions used by the different studies that measured the reference stellar parameters. But 
as argued by \cite{dolowfe},  offsets are  also due to systematics in the model spectra. 
We tested if the systematic uncertainties change as a function of S/N by adding noise to the reference spectra, but found consistent results for lower S/N.
We added the standard deviations in quadrature to our statistical uncertainties to account for the systematics of the model spectra. 

The systematic uncertainty values we obtained are lower than the systematic uncertainties found by \cite{dolowfe}. The reason for this is twofold. \cite{dolowfe} used the MARCS grid \citep{gustafssonmarcs08} and not the PHOENIX grid \citep{husserphoenix}, and they used the IRTF SPEX library \citep{irtf} as reference stars. The IRTF stellar library has a  spectral resolution of $R$ $\approx$ 2\,000. The MARCS grid was computed with  different stellar abundances and atomic line lists than the PHOENIX synthetic spectra. We tested the MARCS grid for our reference stars and found $\langle \Delta\,\teff \rangle$\,=\,$-$38\,K, $\sigma_{\Delta \teff}$\,=\,268\,K, $\langle\Delta{\mh} \rangle$\,=\,$-$0.33\,dex, $\sigma_{\Delta \mh}$\,=\,0.25\,dex,  $\langle\Delta{\logg} \rangle$\,=\,1.2\,dex, $\sigma_{\Delta \logg}$\,=\,0.7\,dex. There appears to be a systematic offset to higher values of $\logg$ and lower values of $\mh$ compared to the reference values. Hence, we decided to use the PHOENIX synthetic spectra. 
However, there are also some systematic offsets when we use the PHOENIX grid.  The effective temperature tends to be overestimated at $\teff$\,$\lesssim$\,4\,000\,K (by up to 360\,K, see lower left panel of Fig. \ref{fig:refcomp}), but rather underestimated at  $\teff$\,$\gtrsim$\,4\,500\,K  (by up to 480\,K). On the other hand, the results for $\logg$ are rather too high at $\logg$\,$\ga$\,2.5\,dex, but too low  at $\logg$\,$\la$\,1\,dex. The metallicities of most spectra are in good agreement with the reference metallicities. We   show two reference star fits in Fig.~\ref{fig:refspec}. We plotted the  GNIRS spectrum of HD206067 (bottom, black). For this star, our results (red) are in good agreement with the reference stellar parameters (blue). 
There is one  outlier for the metallicity results. The fit of the NIFS spectrum of the K4~III star HD121447 indicated a very high metallicity ($\mh$\,$\ga$\,+0.72\,$\pm$\,0.05\,dex), though the reference iron abundance is roughly solar ($[Fe/H]$\,=\,+0.05\,dex; \citealt{smith84}). The observed spectrum from \cite{winge}, the best-fitting model spectrum, and a model spectrum with the reference stellar parameters are shown in Fig. \ref{fig:refspec} (top). By eye, one can hardly decide which model spectrum fits the data better, however, some lines (marked with black dot--dashed vertical lines) are recovered better by our best fit than by the model with the reference values. There are some lines  in the observed spectrum that are missing or only weak in the  model spectra. Some of these lines are marked as green dashed vertical lines  in Fig. \ref{fig:refspec}. 
We conclude that fits, which obtain high metallicities $\mh$\,$\ga$\,+0.5\,dex, must be considered suspect. We note that we  did not test our method on stars with high metallicities beyond $\mh$\,=\,+0.3\,dex. 
 As can be seen from Fig.~\ref{fig:refcomp}, the fitted results in $\teff$ appear to be very robust. A large discrepancy in the fitted metallicity or surface gravity still produces good results in $\teff$.

\subsubsection{Total uncertainties}
\begin{figure}
  \centering  
  \includegraphics[width=0.99\columnwidth]{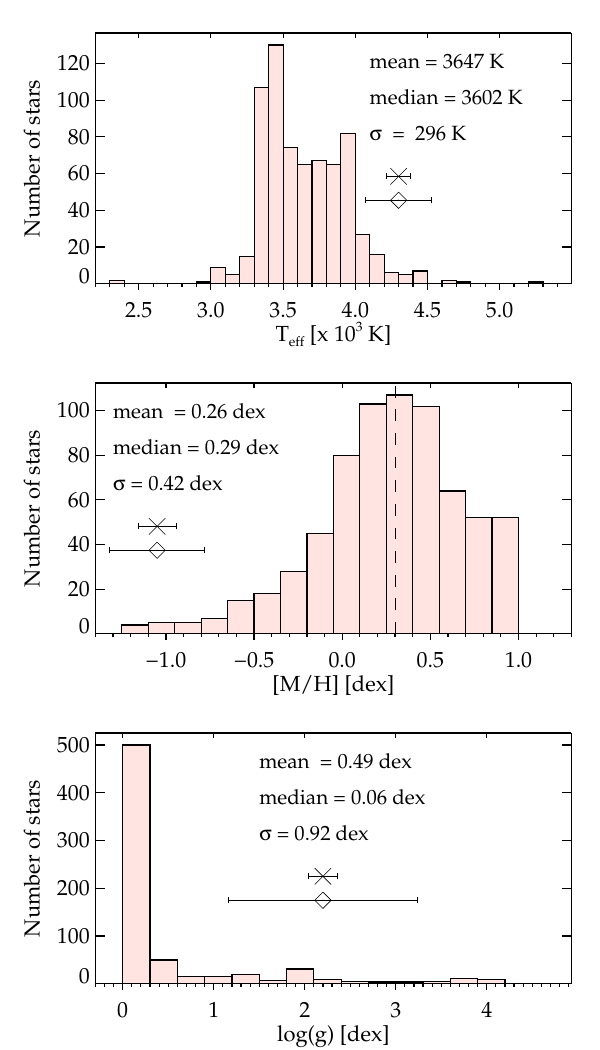}
 \caption{Histograms of the stellar parameters for \goodstarmember\space stars with $(H-K_S)_0$\,=\,[$-$0.5\,mag; 0.8\,mag], from top to bottom: effective temperature $\teff$,  metallicity $\mh$, surface gravity $\logg$. The error bars denote the mean statistical  uncertainty (cross symbol), and mean total uncertainty (diamond symbol) of the measurements.}
 \label{fig:meanhist}
\end{figure}

We took different uncertainties into account and added them in quadrature. These uncertainties are (1) the statistical uncertainties, either the 1$\sigma$ fitting uncertainties or the standard deviation of several exposures of the same star and (2) systematic uncertainties of the synthetic model spectra. 
The systematic uncertainties are dominating the total uncertainty for most of the stars. 
The mean total uncertainties of our sample of  stars are $\sigma_{\teff}$\,=\,230\,K,  $\sigma_{\mh}$\,=\,0.27\,dex, $\sigma_{\logg}$\,=\,1.04\,dex and $\sigma_{v_z}$\,=\,2.9\,km\,s$^{-1}$. The mean statistical and total uncertainties are plotted as error bars in Fig.~\ref{fig:meanhist}.

\section{Results}
\label{sec:sec4}

Our data set contains \ltstar\space stars in the central 64.9\,arcsec $\times$ 43.3\,arcsec of the Milky Way, for which we fitted stellar parameters. We excluded  several stars due to low signal-to-noise ratio or  large uncertainties (Section \ref{sec:sigmult}). Stars with a rather blue colour, or no colour information, may be  foreground  stars, and  not belong to the Milky Way nuclear star cluster. We  excluded those \nocol\space stars, which leaves \goodstarmbg\space stars. A further \bgcut\space stars have rather red colours, they are probably  locally embedded in dust, but might also be background stars. 
Fig.~\ref{fig:meanhist} illustrates the distribution of the stellar parameter results for \goodstarmember\space stars, which are probably cluster members, with extinction corrected colour $(H-K_S)_0$\,=\,[$-$0.5\,mag, 0.8\,mag]. The mean statistical and total uncertainties for the stellar parameter measurements are shown as error bars.

\subsection{Effective temperatures}
As expected, the late-type stars are mostly cool, with a mean temperature $\langle \teff\rangle$\,$\approx$\,3\,650\,K (upper panel of Fig.~\ref{fig:meanhist}). This means the stars are mostly late K-type stars or M-type stars \citep{allen}. As shown in Section~\ref{sec:fitlib}, we rather overestimate low temperatures and underestimate high temperatures. This indicates that the  distribution of $\teff$ might be too narrow. To estimate this effect, we made a linear fit to the $\teff$ measurements of reference stars (red dashed line in upper left panel of Fig. \ref{fig:refcomp}). We shifted our $\teff$ measurements using this relation, which broadens the  $\teff$ distribution by 77\,K. 
Only one star (Id 2164) has a temperature higher than 5\,000\,K, in agreement with a G-type star. Also the CO equivalent width is rather low (4.32\,$\AA$), which confirms the higher effective temperature.

\subsection{Metallicities}
\label{sec:highmet}

The mean metallicity is  $\langle \mh \rangle$\,=\,+0.26\,dex, this means  that most stars are metal-rich. The standard deviation of $\mh$ is 0.42\,dex. 
 The metallicity distribution is shown on the middle panel  of Fig. \ref{fig:meanhist} for \goodstarmember\space likely cluster member stars.  
We found  several metal-poor stars. Our data contain  \lowmetmemberbgzero\space stars with $\mh$\,\textless\,0\,dex, from which \lowmetmemberbgmofive\space stars have $\mh$\,$\leq$\,$-$0.5\,dex and \lowmetmemberbgmone\space stars even  $\mh$\,$\leq$\,$-$1.0\,dex. One of the metal-poor stars with $\mh$\,$\leq$\,$-$0.5\,dex has a rather red colour [$(H-K_S)_0$\,=\,0.84\,mag], it  might either  be locally embedded or a background star. Our data set contains (5.2$^{+6.0}_{-3.1}$) per cent low-metallicity stars (i.e. $\mh$\,$\leq$\,$-$0.5\,dex), and (22.8$^{+24.5}_{-12.6}$) per cent subsolar metallicity stars ($\mh$\,$\leq$\,0.0\,dex).

Several stars are close to the boundary metallicity of the grid at $\mh$\,=\,+1\,dex. This would mean that these stars are super-metal-rich. However, there is probably a problem with the fit or the spectral grid. We only tested the metallicity measurement on reference stars with $\mh \la$ +0.3\,dex. We indicated this metallicity as a vertical dashed line in the middle panel of Fig. \ref{fig:meanhist}. We cannot say whether our method works for stars with a higher metallicity. Further, we showed in Section~\ref{sec:fitlib} that the metallicity of a star with  $\mh$\,\textgreater\,0\,dex was overestimated by nearly 0.7\,dex.  We conclude that stars with a metallicity $\mh \ga$ +0.3\,dex are probably metal-rich ($\mh \ga$ 0\,dex), however, with our method and models at moderate  spectral resolution we cannot determine the metallicity to a higher accuracy. 
Fig.~\ref{fig:fourspec} shows three spectra with different metallicities $\mh$\,=\,$-$0.89,  +0.25 and +0.69\,dex. Their effective temperatures and surface gravities are very similar. 
These spectra have signal-to-noise values of S/N$_\text{fit}$ $\geq$ 30.
For all three spectra,  the Na doublet and Ca triplet  lines (marked as grey shaded area) are deeper than  in the best-fitting spectra. For the metal-poor star,  many of the other lines are fitted reasonably well, e.g. the  Fe lines at 21\,290, 21\,782, 21\,901, 22\,266, or 22\,399\,$\AA$, which are blended with Ti, Si, or Sc. The lines at  21\,290, 21\,782, and 21\,901\,$\AA$ are also fitted well for the spectrum with $\mh$ = +0.25\,dex. However, in particular the  spectrum with the highest metallicity $\mh$ = +0.69\,dex has many lines that are deeper than the best-fitting model spectrum. 
This confirms that the stars with a best-fitting metallicity $\mh$\textgreater +0.3\,dex are definitely very interesting targets to be followed-up with high-resolution spectroscopy. At this point, it would be premature to claim that these stars are indeed super-metal-rich.

\subsection{Surface gravities}
The surface gravity is mostly low, in agreement with cool giant [$\logg$\textless 3.0\,dex] and supergiant [$\logg$\textless 1.5\,dex] stars (lower panel of Fig. \ref{fig:meanhist}). For most stars we obtained a value of $\logg$ close to zero, at the edge of the PHOENIX spectral grid. This suggests that the value of $\logg$ might be even negative, as for  M-type supergiants. However, we showed in Section~\ref{sec:fitlib} that $\logg$ is rather underestimated  at low values of $\logg$, and the systematic uncertainty is 1\,dex. Therefore, we consider the results of $\logg$ as highly uncertain.  

We fitted the surface gravity together with the other stellar parameters but treated it mostly as a nuisance parameter, rather than an actual measurement.
We did this to avoid any biases that may be introduced by fixing $\logg$ to a wrong value.  
As we showed in Section \ref{sec:fitlib}, fitting a discrepant value in $\logg$ does not directly translate into a wrong value for the effective temperature or metallicity. Some of the most discrepant results for $\logg$ give very good results  for $\teff$ and $\mh$. 

\subsection{Radial velocities}
\label{sec:rvlt}
We fitted the radial velocities together with the stellar parameters. They range from $-$268.8 to +313.6\,km\,s$^{-1}$. The mean velocity of  \goodstarmbg\space stars with $(H-K_S)_0$\,\textgreater\,$-$0.5\,mag is  $\langle v_z \rangle$\,=\,+7.6\,$\pm$\,3.6\,km\,s$^{-1}$. The disagreement from  zero is probably due to the asymmetric spatial distribution of stars in our data set, with 410 stars in the Galactic East, and 295 stars in the Galactic West.  
The velocity dispersion  $\langle \sigma_{z} \rangle$\,=\,96.2\,$\pm$\,2.6\,km\,s$^{-1}$. We computed the mean velocity and velocity dispersion  with the maximum likelihood approach \citep{pryor}, which takes the individual velocity uncertainties into account. 

We also considered the Galactic East and Galactic West separately. We obtained  for the  410 stars in the Galactic East $\langle v_{z} \rangle$\,=\,+24.6\,$\pm$\,4.8\,km\,s$^{-1}$,  $\langle \sigma_{z} \rangle$\,=\,96.4\,$\pm$\,3.4\,km\,s$^{-1}$, and $\langle v_{z} \rangle$\,=\,$-$16.2\,$\pm$\,5.3\,km\,s$^{-1}$,  $\langle \sigma_{z} \rangle$\,=\,90.7\,$\pm$\,3.8\,km\,s$^{-1}$ for the 295 stars in the Galactic West. The mean radial velocity  confirms the rotation of the Milky Way nuclear star cluster by approximately 20\,km\,s$^{-1}$ within the central 1.2\,pc (30\,arcsec), as found by \citet{mcginn89} and also \cite{isaacanja} from a different data set. 

\begin{figure*}
      \includegraphics[width=0.95\textwidth]{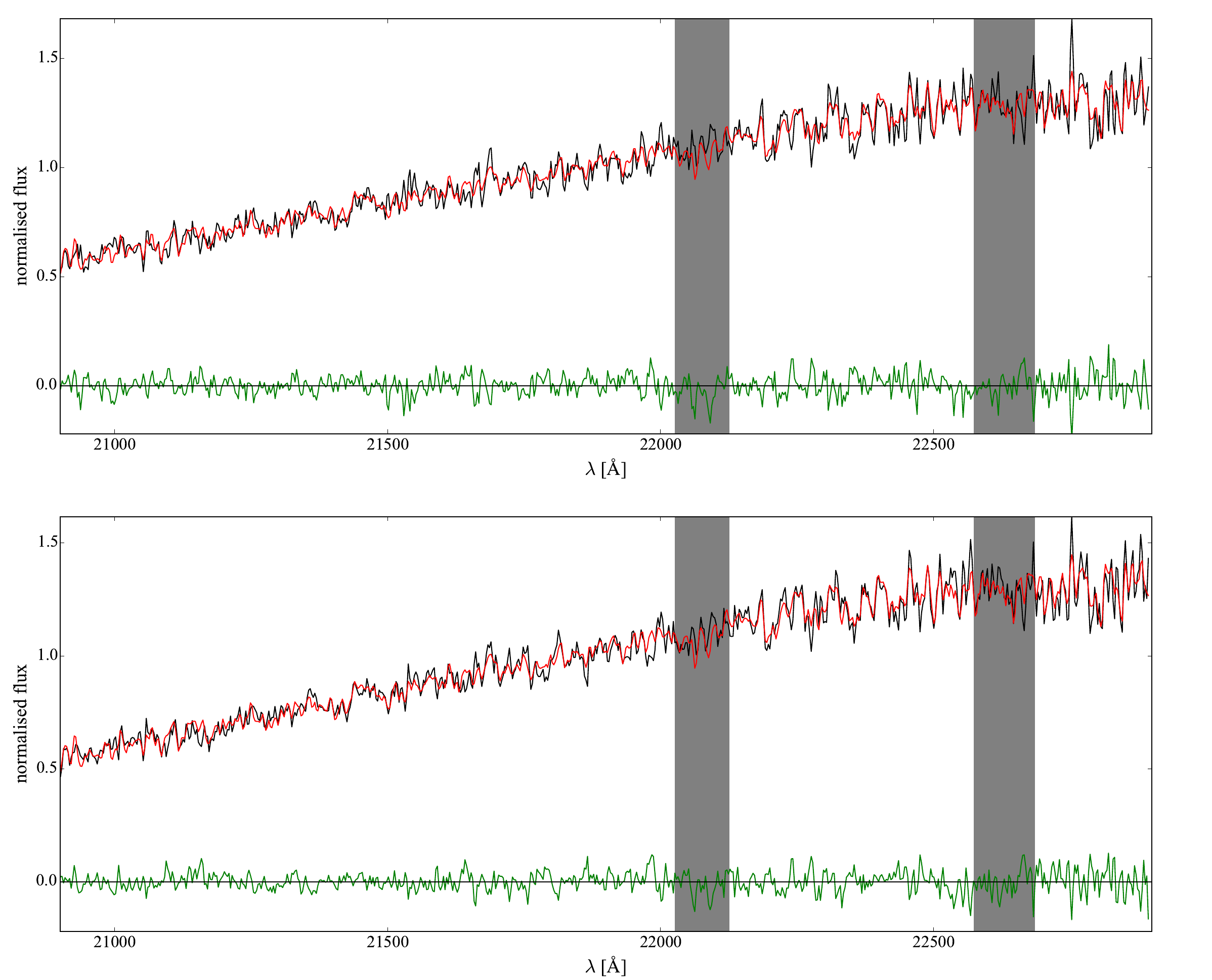}
  \caption{Spectra of the long-period variable star Id 649 in black. We fit the two exposures separately, the best fit (red) is obtained at  $\teff$\,=\,2\,300\,K, $\mh$\,=\,+0.99\,dex, $\logg$\,=\,5\,dex. Green circles denote the fit residuals. The flux of the spectrum increases towards longer wavelengths.  The fits have $\chi^2_\mathrm{red}$=11.6 and 14.2, and S/N$_\text{fit}$=20.6 and 22.5.}	
 \label{fig:cool649}
\end{figure*}

\subsection{Long-period variable stars}
\label{sec:lpv}
Our data contain six stars with $\teff$\,\textless\,3\,000\,K, five of them have an effective temperature close to the edge of the PHOENIX grid, at $\teff$\,=\,2\,300\,K. 
We checked those  stars in the SIMBAD data base \citep{simbad} and found that four of the stars were classified as long-period variables or Mira stars (Id 158, 221, 25349, 25389) by \cite{2003reid} and \cite{matsunaga09}.  Such stars are mostly M-type giants or supergiants, with periods of few to several hundred days. However, long-period variables can also be carbon-rich  (C-type stars), or zirconium-rich (S-type stars), and Mira stars can be oxygen-rich or  carbon-rich  \citep{allen}. 

The six stars are rather bright, with observed magnitudes  ranging from $K_S$ = 9 to 12.3\,mag. This also indicates that the stars are supergiant and bright giant stars, however,  we obtained $\logg$  values ranging from 1.5 to 5.6\,dex.  We conclude that the PHOENIX spectral grid is not suitable to obtain the surface gravity of  C-type, S-type or oxygen-rich  long-period variable stars. 
Many Mira stars suffer from rapid mass-loss. These stars are often embedded in dust, which causes reddening. Indeed, three of these six cool stars have rather red extinction-corrected colours [$(H-K_S)_0$\,=\,0.93--2.12\,mag].
This reddening makes it hard to classify the six cool stars as member  or background stars. 
Further, all six stars have a steeply rising continuum  (slope m\,=\,$\Delta$flux/$\Delta \lambda$\,=\,3.2--4.8, see also Fig.~\ref{fig:cool649}), at least three times steeper than most other stars. This also indicates that the stars are embedded in dust, and our photometric extinction correction was possibly too low.

\subsection{Spatial distribution and kinematics of metal-poor stars}
\label{sec:frac}
\begin{figure}
  \centering  
  \includegraphics[width=0.99\columnwidth]{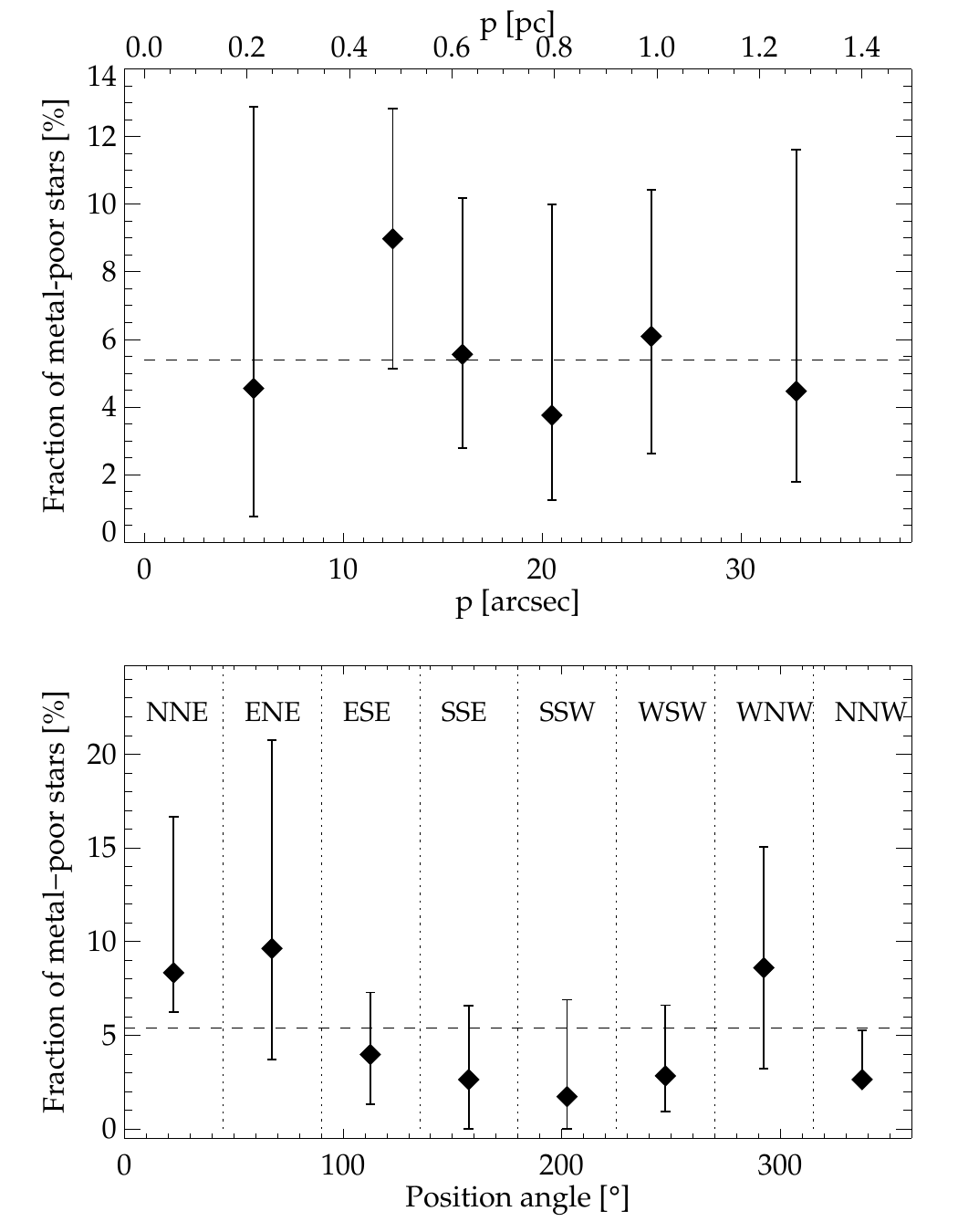}
 \caption{Fraction of metal-poor stars ($\mh$\,$\leq$\,$-$0.5\,dex) as a function of projected distance $p$ from Sgr~A* (upper panel), and as function of the position angle (Galactic East of North) centred on Sgr~A* (lower panel). }
 \label{fig:frac}
\end{figure}

We investigated if the metal-poor stars show any kind of peculiar spatial distribution. For this purpose, we counted the number of metal-poor stars ($\mh$\,$\leq$\,$-$0.5) in circular annuli around Sgr~A*. We chose the bins such that each bin contains at least five metal-poor stars. The fraction of metal-poor stars divided by  the total number of late-type stars is shown in the upper panel of Fig.~\ref{fig:frac}. The uncertainties come from the total uncertainty of $\mh$. We did not find a significant change of the fraction of metal-poor stars in the    range of 0--1.4\,pc projected distance from Sgr~A*. 
 Further, we  divided the data set in eight segments, and computed the fraction of metal-poor  stars with respect to the total number of stars as a function of the position angle. The result is shown in Fig.~\ref{fig:frac} (lower panel). We found that the fraction of metal-poor stars is highest in the Galactic  North East, and lower in the Galactic South. But again, the uncertainties are too high for this trend to be significant. 

\begin{figure}
  \centering  
  \includegraphics[width=0.99\columnwidth]{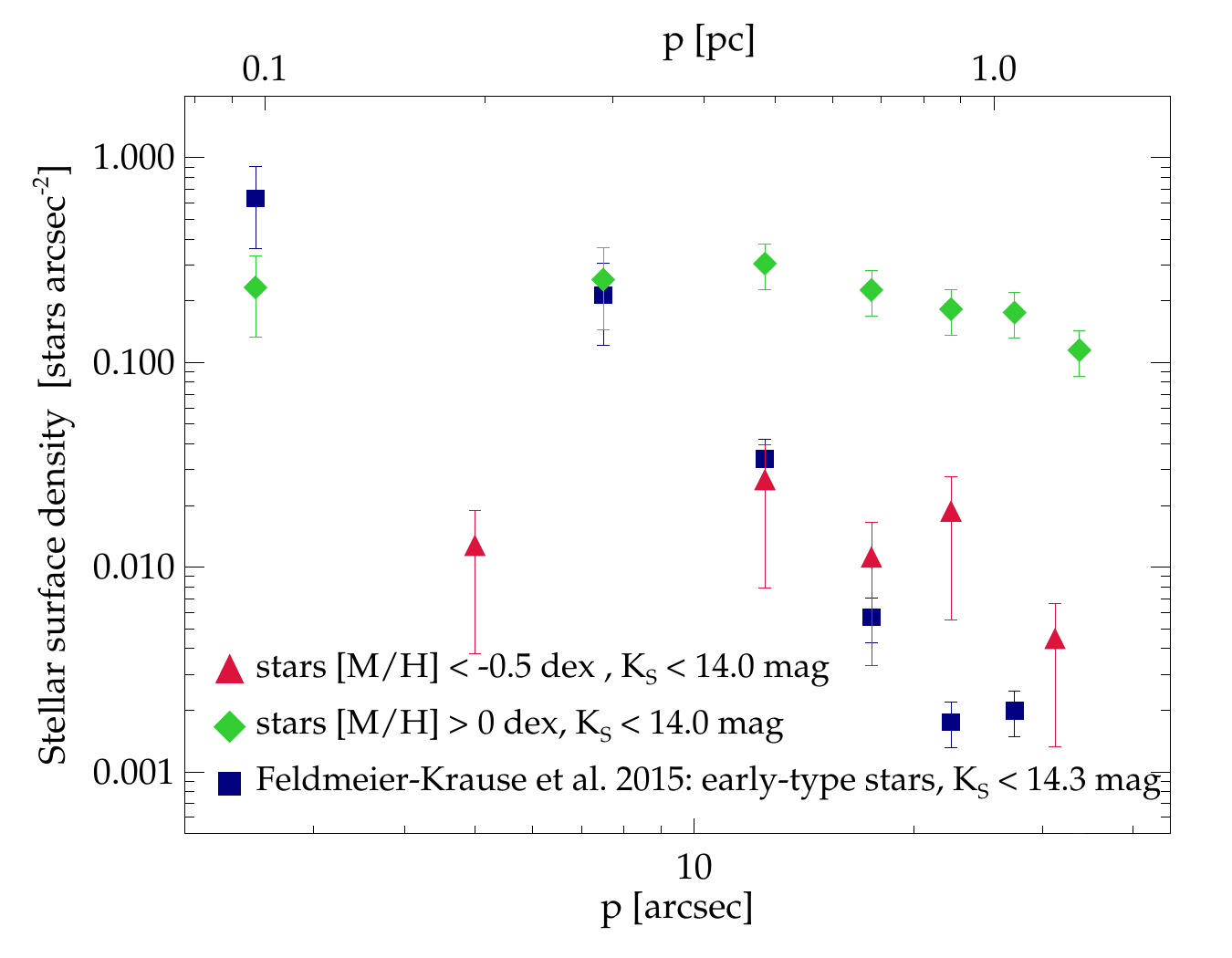}
 \caption{Completeness corrected stellar surface number density profile for stars brighter than $K_S$ = 14.0\,mag. Red triangles denote metal-poor stars ($\mh$\,$\leq$\,$-$0.5\,dex), green diamond symbols denote supersolar metallicity stars ($\mh$\,$\geq$\,0.0\,dex). Blue square symbols illustrate the distribution of early-type stars. }
 \label{fig:ndensmet}
\end{figure}

We show a completeness corrected surface number density profile of metal-poor stars (red triangles) and metal-rich stars ($\mh$\,$\geq$\,0.0\,dex, green diamond symbols) in Fig.~\ref{fig:ndensmet}. We used the same method as in \cite{kmoset} to construct the profile, i.e. we corrected the number counts in different magnitude and radial bins for completeness, and counted the stars in rings around Sgr~A*. The surface number density of early-type stars is denoted with blue square symbols. The decrease of the  early-type star density to larger radii is steep, over more than two orders of magnitude. The density of stars with $\mh$\,$\geq$\,0.0\,dex and $K_S$\,\textless\,14\,mag is nearly constant, and decreases only by a factor $\la$3. There is much scatter in the surface number density of metal-poor stars. 

\begin{figure}
  \centering  
  \includegraphics[width=0.99\columnwidth]{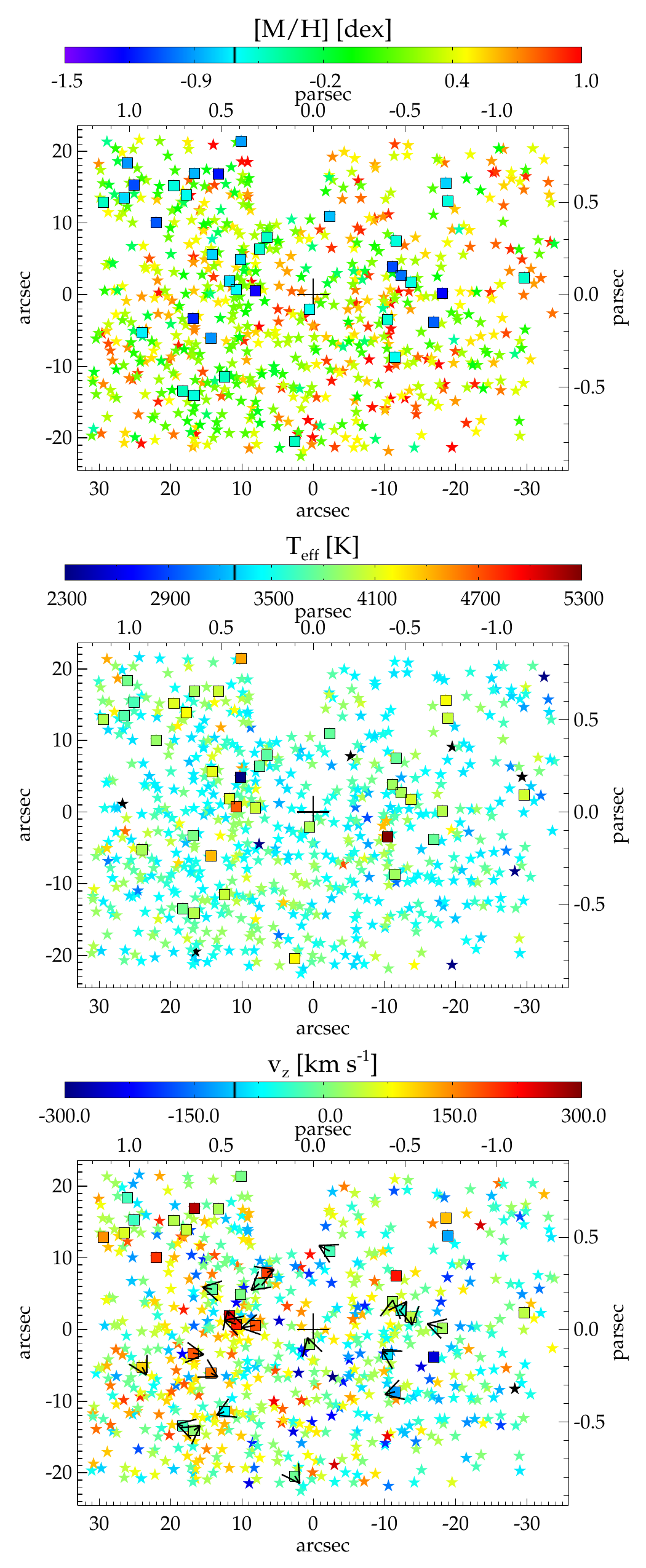}
   \caption{Spatial distribution of \goodstarmbg\space stars with $(H-K_S)_0$\,\textgreater\,$-$0.5\,mag in  offset coordinates  from the central supermassive black hole, Sgr A*, illustrated as black cross. Galactic North is up. The colours denote the metallicity $\mh$ (upper panel), effective temperature $\teff$ (middle panel) and radial velocity $v_z$ (lower panel). The square symbols highlight  metal-poor stars with $\mh$\,$\leq$\,$-$0.5\,dex. Black arrows denote proper motions of metal-poor stars. }
 \label{fig:spatialcol}
\end{figure}

We also illustrate the spatial distribution of the \goodstarmbg\space stars with $(H-K_S)_0$\,\textgreater\,$-$0.5\,mag in Fig.~\ref{fig:spatialcol}. The upper panel is colour-coded with the metallicity $\mh$, and metal-poor stars  are highlighted as square symbols. The middle panel is colour-coded with the effective temperature. Metal-poor stars are rather hotter than the other stars, with $\langle \teff \rangle$\,=\,3\,980\,K.   Only one metal-poor star is a long-period variable star with a low temperature of 2\,300\,K. We already discussed in Section~\ref{sec:lpv} that long-period variable stars can have an unusual chemical composition, and therefore the results are uncertain.

The radial velocity $v_z$ is illustrated in the lower panel of Fig.~\ref{fig:spatialcol}. The metal-poor stars also rotate around Sgr~A*, and their mean velocity and velocity dispersion are 
$\langle v_{z} \rangle$\,=\,+76.8\,km\,s$^{-1}$,  $\sigma_{z}$\,=\,101.9\,km\,s$^{-1}$ in the Galactic East,     $\langle v_{z} \rangle$\,=\,$-$15.2\,km\,s$^{-1}$,  $\sigma_{z}$\,=\,124.3\,km\,s$^{-1}$ in the Galactic West,  and   $\langle v_{z} \rangle$\,=\,+47.0\,km\,s$^{-1}$,  $\sigma_{z}$\,=\,116.4\,km\,s$^{-1}$ in the entire field. The  velocity dispersion is similar to the velocity dispersion  of all stars ($\sigma_{z}$\,=\,94.6\,$\pm$\,2.6\,km\,s$^{-1}$, Section \ref{sec:rvlt}).
We also matched the stars with the proper motion data of \cite{Rainerpm09}, and found 400 matches. The proper motions  of the metal-poor stars are plotted as black arrows in Fig.~\ref{fig:spatialcol}. The average velocity dispersion over  two dimensions for 21 metal-poor stars with proper motions ($\sigma_{2d}$\,=\,89.8\,km\,s$^{-1}$) is in agreement with the velocity dispersion of the other stars ($\sigma_{2d}$\,=\,100.3\,$\pm$\,3.7\,km\,s$^{-1}$). As for the radial velocities,  the absolute two-dimensional velocity of the metal-poor stars ($\langle v_{2d} \rangle$\,=\,126.3\,km\,s$^{-1}$)  is in agreement with  the two-dimensional velocity of the other 379 stars  ($\langle v_{2d} \rangle$\,=\,127.5\,$\pm$\,3.3\,km\,s$^{-1}$). Altogether the kinematics of the metal-poor stars are not significantly distinct from the kinematics of the other stars.

\subsection{Influence of radial velocity  shift}
\label{sec:shiftrv}
Depending on the radial velocity of a star, a slightly different wavelength region is fitted. In this section, we test if this has a strong effect.
We fitted the spectra as observed in the fixed wavelength interval $\lambda$\,=\,[20\,900\,$\AA$, 22\,900\,$\AA$]. But since the stars have velocities in the range of $v_z$\,=\,[$-$269\,km\,s$^{-1}$; 314\,km\,s$^{-1}$], some spectra are shifted with respect to the rest wavelength by up to 22\,$\AA$.  This means we fit the stars in different rest wavelength intervals, and some lines at the edge of $\lambda$\,=\,[20\,900\,$\AA$, 22\,900\,$\AA$] may be part of the fit or not, depending on the stellar radial velocity. To test the magnitude of this effect, we shifted 50 spectra to the rest wavelength, and fitted the spectra at slightly shifted wavelength regions, at $\lambda$\,=\,[20\,900\,$\AA$, 22\,900\,$\AA$], $\lambda$\,=\,[20\,920\,$\AA$, 22\,920\,$\AA$], and $\lambda$\,=\,[20\,880\,$\AA$, 22\,880\,$\AA$]. The offset of $\pm$20\,$\AA$ corresponds to $\pm$270\,km\,s$^{-1}$. 
For these 50 spectra we found that the median of the stellar parameter change is  $\Delta \teff $\,=\,3\,K, $\Delta {\mh}$\,=\,+0.01\,dex, $\Delta {\logg}$\,=\,+0.01\,dex, $\Delta v_{z}$\,=\,0.3\,km\,s$^{-1}$ with  standard deviations of $\sigma_{\Delta \teff}$\,=\,44\,K,  $\sigma_{\Delta \mh}$\,=\,0.1\,dex,  $\sigma_{\Delta \logg}$\,=\,0.6\,dex, and $\sigma_{\Delta v_z}$\,=\,0.7\,km\,s$^{-1}$. 
This means that a shift in the radial velocity does not bias the results of the stellar parameters, though it can change the results. The difference is usually less than the uncertainties. 

\subsection{Influence of spectral resolution}
\label{sec:shiftspecr}
As we showed in Section \ref{sec:lsf}, the spectral resolution depends on the used IFU. For the spectral fits,  we took the different spectral resolutions into account. In this section, we test if a lower spectral resolution obtains consistent results with a higher spectral resolution, and how a mismatch of the data and model spectral resolution affects the results. 

We used the spectral library and convolved the spectra to the minimum and maximum spectral resolution  of the KMOS data ($R$\,=\,3\,300 and 4\,660).  We fitted the spectra with $R$ = 3\,300 using model spectra at $R$ = 4\,660, and the spectra with  $R$ = 4\,660 using model spectra at $R$ = 3\,300. The results of $\teff$, $\logg$ and $v_z$ change only slightly, and there are no systematic offsets. Only the metallicity is affected. In particular, when fitting the data with models that have a higher spectral resolution, the metallicity is shifted by $\langle \Delta\mh \rangle$ = +0.1\,dex with $\sigma_{\Delta\mh}$ = 0.06\,dex. On the other hand, fitting data using model spectra that have a lower spectral resolution, the metallicity is shifted by $\langle \Delta\mh \rangle$ = $-$0.09\,dex with $\sigma_{\Delta\mh}$ = 0.05\,dex.

We repeated the fits using model spectra convolved to the correct respective spectral resolution. The mean differences and standard deviations at the different spectral resolutions are small compared to the uncertainties. We conclude that a variation of the spectral resolution  does not bias our  results when the correct spectral resolution is adopted for each spectrum. 

\section{Discussion}
\label{sec:sec5}
In this section, we discuss our results of the stellar metallicities, and compare them with other metallicity measurements. We argue that the stars in our data set are likely members of the nuclear star cluster. From the stellar $K$-band magnitude, we infer the luminosity classes and conclude that the metal-poor stars are consistent with globular cluster stellar populations.  We finally discuss implications for the formation of the Milky Way nuclear star cluster.

\subsection{Detection of stars with low metallicity $\mh$\,$\la$\,$-$0.5\,dex}
\label{sec:discmet}
We measured the metallicities of \goodstarmbg\space stars in the near-infrared $K$ band with medium spectral resolution.  We detected \lowmetmemberbgmofive\space stars with $\mh$\,$\la$\,$-$0.5\,dex, this is  5.2 per cent of the stars in our sample. Their mean magnitude $K_{S,0}$\,=\,13.18\,mag is similar to the mean magnitude of the stars with $\mh$\,\textgreater\,$-$0.5\, dex ($K_{S,0}$\,=\,13.04\,mag). 
We could not calibrate the metallicity measurements with  metal-poor library stars with $\mh$\,\textless\,$-$0.5\,dex. Therefore, we could not test if we can differentiate between stars with $\mh$\,=\,$-$0.5\,dex and $\mh$\,=\,$-$1.0\,dex.  But we showed in Section~\ref{sec:fitlib} that our method is able to differentiate metal-poor  ($\mh$\,\textless\,$-$0.3\,dex) from metal-rich stars ($\mh$\,\textgreater\,+0.0\,dex). 
 
Our metallicity results are in agreement with the results of   \cite{dolowfe}, who measured the metallicities of 83 Galactic Centre stars.
\cite{dolowfe} used the same fitting procedure as this work, but their data have a higher spatial and  spectral resolution  ($R$\,=\,5\,400). Further, they used the MARCS grid \citep{gustafssonmarcs08}. We matched their 83 stars with our data set. To obtain a match, the distance of two stars has to be  less than  0.3\,arcsec, and the difference in $K_S$-band magnitude  not more  than 0.4\,mag. With these conditions,  we have 29 stars in common. 
The temperatures and metallicities are in reasonable agreement, with $\langle \Delta \teff \rangle$\,=\,$-$2.6\,K, $\sigma_{\Delta \teff}$\,=\,148\,K,  $\langle \Delta \mh \rangle$\,=\,$-$0.32\,dex, $\sigma_{\Delta\mh}$\,=\,0.35\,dex. The values of $\logg$  have a larger disagreement than expected from fitting the spectral library in Section \ref{sec:fitlib}, with  $\langle \Delta{\logg} \rangle$\,=\,$-$2.6\,dex, $\sigma_{\Delta \logg}$\,=\,0.9\,dex. This is caused by the low surface gravities obtained with  the PHOENIX grid,  and the tendency of the MARCS grid to higher surface gravities \citep[see fig. 2 of][]{dolowfe}. The radial velocity is also offset by $\langle \Delta v_z \rangle$\,=\,28.9\,km\,s$^{-1}$, $\sigma_{v_z}$\,=\,7.5\,km\,s$^{-1}$, but this is simply due to the fact that \cite{dolowfe} did not shift their spectra to the local standard of rest, as we did. This is a shift of less than 2\,$\AA$, and should therefore not affect the results severely (see Section \ref{sec:shiftrv}). 

\subsection{Note of caution on the measurement of high metallicities $\mh$\,$\ga$\,$+$0.5\,dex }
Our results suggest that most stars are metal-rich, i.e., $\mh$\,\textgreater\,0.0\,dex. For more than 75 per cent of the  stars we  obtain a metallicity $\mh$\,\textgreater\,0.0\,dex, and for about 28 per cent even $\mh$\,\textgreater\,+0.5\,dex. The measurements  of high metallicities have to be regarded with caution. Our metallicity measurements are differential and may not be correct in absolute values.

As \cite{dolowfe} pointed out, there might be systematic effects due to the use of medium-resolution spectroscopy. 
Differences between metal-rich and solar-metallicity spectra are not as pronounced as differences between metal-poor and solar-metallicity spectra. 
We  showed  in Section~\ref{sec:fitlib} on the fit of a reference star that  the metallicity can be overestimated by almost 0.7\,dex. Fig.~\ref{fig:refspec} illustrates that  several lines of the  top spectrum (black) are deeper than the lines of the best-fitting model spectrum (red). The model spectrum with the reference stellar paramters (blue) of \cite{smith84} is a worse fit. 
A similar  effect can be seen in the KMOS data.
In Fig.~\ref{fig:fourspec},  we showed the spectrum of a star with supposedly high metallicity ($\mh$\,=\,+0.69\,dex, bottom spectrum). Although the model spectrum (red) has a high metallicity, several lines of the observed spectrum (black) are still deeper and not fitted well. It might be that the used model spectra are not ideal to fit red giants with high metallicities at medium spectral resolution. Maybe the assumption of solar $\alpha$-abundances causes problems at high metallicities.  
Alternatively, the line strengths of strong lines in the model spectra may be too low if   the microturbulence assumed by  \cite{husserphoenix} was too low. 
Further, the synthetic spectra were computed assuming local thermodynamic equilibrium (LTE), which may affect our results \citep{2012MNRAS.427...50L,2014dapb.book..169B}. However, \cite{2012MNRAS.427...50L} showed that non-LTE effects are most important for extremely metal-poor stars ([$Fe/H$]\textless $-$3\,dex) and hot ($\teff$\,\textgreater 5\,000\,K) giants, and can lead to an underestimation of the Fe abundance from optical spectra. If the same is true for metallicities in the near-infrared, our results for the cool, metal-rich giants  are  unaffected by non-LTE effects. 
 The \textit{Gaia}-ESO Survey and the APOGEE measured metallicities  for several thousand stars of the Milky Way \citep{2014A&A...572A..33M,2015AJ....150..148H}. They  found some stars with high metallicities $\mh$\,$\approx$\,+0.5\,dex. However, their spectra are at shorter wavelengths than our data, and we cannot compare their spectra of metal-rich stars to our data.
 In summary, we cannot rule out that some of the stars in our data set have high metallicities.  But since the spectra of the alleged super-metal-rich stars are not fitted well,  we do not claim the detection of such stars. 
 
 Nevertheless, we conclude that most of the stars in our data set are metal-rich ($\mh$\textgreater\,+0.0\,dex).
 Studies of the Milky Way bulge and disc showed that the metallicity increases towards the centre \citep[e.g.][]{2008A&A...486..177Z,2015ApJ...808..132H,2016PASA...33...22N}, with the most metal-rich stars concentrated to the plane. Therefore, it is not surprising to find  metal-rich stars in the Galactic Centre. 
To verify the metallicity measurements, observations with higher spectral resolution are  required. A subsample of the metal-poor stars and some metal-rich stars observed with high-resolution spectroscopy would be useful to calibrate our measurements, and to measure abundance ratios, such as  [$Na/Fe$], and [$Ca/Fe$]. 
 But also the applied tools should be improved. As the Galactic Centre stars seem to be enhanced in some elements,   synthetic spectra with higher element abundances are probably more suitable to fit the spectra. At the moment, PHOENIX spectra with $[\alpha/Fe]$\,\textgreater\,0\,dex are only available for $\mh$\,=\,0.0\,dex. We plan to implement different $\alpha$-abundances to the spectral fit in the future. In addition, a larger spectral library in the $K$ band at high to medium spectral resolution will be useful. In the near future, the X-SHOOTER spectral library \citep{2014A&A...565A.117C} will provide  near-infrared spectra, which  will be very useful to   study the Milky Way nuclear star cluster.

\subsection{Metallicity distribution}

The metallicity distribution is important to derive the fraction of metal-poor and metal-rich stars. This information is essential  to reconstruct  the star formation history. 
We obtained a  mean metallicity   $\langle \mh \rangle$\,=\,+0.26\,dex. The standard deviation of $\mh$ is 0.42\,dex. 

Other studies obtained similar results for the metallicity distribution in the Galactic Centre. \cite{dolowfe} found a mean value of $\langle $\mh$ \rangle$\,=\,+0.4\,dex and a standard deviation of 0.4\,dex for a sample of 83 stars.  
Further out, in the inner Galactic bulge, \cite{schultheis15} measured the metallicities of 33 stars and obtained a mean metallicity $\langle \mh \rangle$\,=\,$+$0.4\,dex with a dispersion of 0.55\,dex. They also found eight low-metallicity stars with $\mh$\,$\approx$\,$-$1.0\,dex and enhanced $\alpha$-element abundance. 
Our metallicity measurements are in agreement. Due to the larger number of stars in our sample, we were able to find a continuous metallicity distribution.  The metallicity distribution is rather smooth, and we do not see signs of a second peak at $\mh$\,=\,$-$1.0\,dex, as indicated by the smaller samples of \cite{dolowfe} and \cite{schultheis15}. However,  a smaller second metallicity peak may be undetectable given our measurement uncertainties. 
Our metallicity distribution has a  negative skewness, which was also found for the  metallicity distribution of the inner Galactic disc using APOGEE data \citep{2015ApJ...808..132H}. 
However, we note that we did not measure systematic uncertainties for metallicities $\mh$\,\textless\,--0.5\,dex and $\mh$\,\textgreater\,+0.3\,dex   on reference stars, but assumed the same systematic uncertainties as  in the range $-$0.5\,dex\,\textless\,$\mh$\,\textless\,+0.3\,dex. The  shape of the metallicity distribution at low ($\mh$\,\textless\,$-$0.5\,dex) and high ($\mh$\,\textgreater\,0.3\,dex) metallicities may be influenced by unknown systematic effects.  Nevertheless, 
the spread of the metallicity distribution by about 0.4\,dex  means that the chemical composition of the stars in the Galactic Centre is inhomogeneous. 


\subsection{Contamination from foreground or background sources}
\label{sec:fgbg}
We checked the membership of a star to the Milky Way nuclear star cluster based on the extinction corrected colours. We identified foreground stars, e.g., from the Galactic bulge or disc,  with their blue colours.  
But as we do not know exactly the extinction along the line of sight, it is possible that some stars of our sample are bar or bulge stars with rather high extinction. However, their number density should be lower than the number density of Galactic Centre stars. \cite{clarkson12} studied a 12\,arcsec $\times$ 12\,arcsec\space field near the Arches cluster, about 26\,pc  in projection away from our field. Their data contain  only one field star with $K$\textless 13.8\,mag, identified by proper motions. This translates to a number density of 0.007 inner bulge stars per arcsec$^2$. In our field of 2\,700\,arcsec$^2$, we estimate the number of inner bulge stars to  approximately 19, but this estimate is based on a small field with only one foreground star in our magnitude range.

We also checked if the kinematics of the stars contain any hints on the cluster membership. For this reason, we matched the stars in our data set with the proper motion data of \cite{Rainerpm09}. We obtained 400 matches, all within the central $p$\,=\,25\,arcsec\,($\sim$0.97\,pc) of the nuclear star cluster. Most of the stars have a lower velocity than the escape velocity, and their kinematics are in agreement with being bound to the nuclear star cluster. There is one exception with a significantly higher velocity than the escape velocity, this is the potential runaway star found by \cite{Rainerpm09}. It has a proper motion velocity of 424\,km\,s$^{-1}$. 
\cite{Rainerpm09} suggested that this star might escape the Milky Way nuclear star cluster.
About 40 per cent of the stars move against the main direction of rotation  of the nuclear star cluster. Several stars in the Galactic East have a negative  velocity $v_z$, and likewise, several stars in the Galactic West have a positive velocity. But this does not mean that these stars do not belong to the nuclear star cluster. They may be on bound counter-rotating tube orbits or on box orbits. 

There might  be some stars from the nuclear stellar disc  with similar colours as the stars of the nuclear star cluster. The nuclear stellar disc and the nuclear star cluster form the nuclear stellar bulge. The   nuclear stellar disc extends over $p$ $\approx$ 120\,pc, and dominates over the nuclear star cluster at $p$\,$\gtrsim$\,30\,pc \citep{launhardt02}.
We are  not able to  distinguish if a star is from the nuclear star cluster or the nuclear stellar disc based on the extinction or kinematics.

\subsection{Luminosity classes and  implications for stellar ages}
The $K_S$-band magnitude can be used to infer the luminosity class of the stars, and estimate their age. 
The extinction corrected magnitudes of the stars in our data set range from $K_{S,0}$\,\textless\,6 to  $K_{S,0}$\,$\approx$\,13.5\,mag, with a median at $K_{S,0}$\,=\,10.4\,mag. 

Stars with magnitudes $K_{S,0}$\,$\lesssim$\,5.7\,mag are red supergiants \citep{blum03}, but supergiants may be as faint as approximately 7.5\,mag \citep{blum96,blum03}.  Red supergiants are massive ($\mathcal{M}$\,\textgreater\,10\,M$_{\sun}$; \citealt{allen}) and young, only a few 10$^7$\,yr old \citep{2011spug.book.....G}.
For five stars   in the range 5.7\,mag \,$\la$\,$K_{S,0}$\,$\la$\,7.5\,mag, one cannot distinguish supergiants from (bright) giant  stars based on the magnitude. \cite{blum03} suggested that it is more likely for the stars to be (bright) giants.

Red giant stars have masses of about 0.3--10\,M$_{\sun}$, 
and ages  $\gtrsim$1\,Gyr. 
The stars with $K_{S,0}$\,\textgreater\,7.5\,mag are probably red giant stars. This means  99 per cent of the stars in our data set are red giants.
The red clump is a subgroup of red giants, their magnitude is  roughly $K_{S,0}$\,=\,12.9\,mag \citep{rainer10}.   Only about 1.5 per cent of the stars in our data set are faint enough for the red clump. 
Another subgroup of red giant stars are the bright asymptotic giant branch (AGB) stars.  
AGB stars have masses in the range 0.5\,M$_{\sun}$\,$\la$\,$\mathcal{M}$\,$\la$\, 10\,M$_{\sun}$ \citep{blum96}.  
They  have two phases, the early AGB (E-AGB) phase, and a later thermally pulsing AGB (TP-AGB) phase.  
During the later phase, material from the core is mixed with the surface material. The six long-period variables (Section~\ref{sec:lpv}) are probably such AGB stars. 
TP-AGB stars are rather bright, with $K_{S,0}$\,$\lesssim$\,8.4\,mag for solar and higher metallicities and ages \textgreater 1\,Gyr \citep[PARSEC evolutionary tracks; ][]{bressan12,2016ApJ...822...73R}. About 50 stars (7\,per cent)  in our sample could be TP-AGB stars, based on their $K_S$-band magnitudes. 
Fainter stars  could be E-AGB stars (6.9\,mag\,$\lesssim$\,$K_{S,0}$\,$\lesssim$\,12\,mag, 93 per cent) or normal red giant branch stars (7.5\,mag\,$\lesssim$\,$K_{S,0}$\,$\lesssim$\,16.4\,mag,  99 per cent). As the time a star spends on the red giant branch is about 40 times longer than the time spent as AGB  star \citep{2011spug.book.....G},  it is more likely for a star to be   on the red giant branch.

The metal-poor stars  in our data set have magnitudes ranging from $K_{S,0}$\,=\,7.15 to 
12.56\,mag and temperatures from 3\,660 to 5\,300\,K, except for one cooler long-period variable star. We compared the location of the stars in a $\teff$--$K_{S,0}$ diagram  with the PARSEC evolutionary tracks \citep{bressan12}.  Using $\mh$\,=\,$-$0.5 and $-$1.0\,dex for the tracks, we find that the metal-poor  stars have ages $\ga$1\,Gyr, and initial masses $\mathcal{M}_{\mathrm{ini}} $\,$\la$\,2\,M$_{\sun}$. Tracks with  age 10\,Gyr and masses $\mathcal{M}_\mathrm{ini}$\,$\la$\,1\,M$_{\sun}$ are also in agreement with the stellar effective temperatures and magnitudes.  This means the stellar ages and masses are consistent with stars in Galactic  globular clusters.

\subsection{Clues to the formation of the  Milky Way nuclear star cluster}
There is a  large spread of metallicities  ranging from metal-poor stars ($\mh$\,$\la$\,$-$1.0\,dex) to  metal-rich stars ($\mh$\,$\ga$\,+0.2\,dex) in the late-type giant star population of the Milky Way nuclear star cluster. This means that the cluster did not form from one homogeneous gas cloud in one single starburst. 

The metal-poor stars in the Milky Way nuclear star cluster may have formed in an early starburst \textit{in situ}, or somewhere else in the Galaxy, and migrated to the centre. The stars might have migrated from the Milky Way bulge to the centre, as the inner bulge also contains some metal-poor stars \citep{schultheis15,2016PASA...33...22N}. Another possibility is that  the stars formed in a star cluster, potentially a globular cluster, that migrated to the centre of the Milky Way.  The Milky Way globular cluster system has a bimodal metallicity distribution, with  about 71 per cent metal-poor ([$Fe/H$]$\la-0.75$\,dex) and 29 per cent metal-rich ($-$0.75$\la$[$Fe/H$]$\la$0.0\,dex) globular clusters  \citep{bica06}. The subsolar metallicity stars in the Milky Way nuclear star cluster have similar metallicities as the `metal-rich' Milky Way globular clusters.

A large fraction of stars in the Milky Way nuclear star cluster is metal-rich ($\mh$\,\textgreater\,0.0\,dex). These stars  cannot originate from the infall of globular clusters to the centre of the Milky Way.   The metal-rich stars must have formed from  enriched gas within the Milky Way, as their metallicities are inconsistent with the observed Milky Way globular cluster metallicities. Metal-rich stars formed either directly in the Galactic Centre, or in an enriched star cluster within the Milky Way that migrated to the Galactic Centre.

The spatial distribution and stellar kinematics  should be able  to discern the \textit{in situ} formation and cluster infall scenario for  the different stellar populations. 
\cite{2015ApJ...799..185A} simulated the \textit{in situ} formation scenario for nuclear star clusters. 
 Depending on where the star formation occurs dominantly, in the centre or in the outskirts of the nuclear star cluster, the cluster would have a positive or negative age gradient. But also the consecutive infall of multiple stellar clusters creates an age and stellar population gradient in the nuclear star cluster \citep{perets14}.
 We investigated the radial distribution of  metal-poor stars, but we were not able to find any significant difference with respect to  metal-rich stars. The number of metal-poor stars  is too low, and the uncertainties are too high to discern the spatial distributions. 
We found indications for a slight overabundance of metal-poor stars in the Galactic North East. If this observation is confirmed,  it would rather suggest a later infall scenario than \textit{in situ} formation. If the old, metal-poor stars have formed \textit{in situ}, they should be distributed isotropically now. However, this assumption needs to be confirmed in three-dimensional simulations of \textit{in situ} nuclear star cluster formation. 
The kinematics of the metal-poor stars are not significantly different from the kinematics of the other stars.
The dynamical evolution of the  different formation scenarios needs to be studied further in simulations.

\section{Conclusions}
We observed the central  4\,pc$^2$ of the Milky Way nuclear star cluster in the $K$ band with the integral-field spectrograph KMOS. We analysed the spectra of more than 700 late-type stars. We found that the equivalent width of the \ion{Na}{I} doublet line region is enhanced with respect to stellar library spectra. Using full-spectral fitting, we derived  effective temperatures and metallicities of \goodstarmbg\space stars of the Milky Way nuclear star cluster. 
Most stars are red K- and M-type giants with temperatures $\teff$\,=\,3\,000--5\,000\,K. The metallicities range from subsolar $\mh$\,\textless\,$-$1.0\,dex to supersolar $\mh$\,\textgreater\,$+$0.3\,dex, with a standard deviation of 0.42\,dex. 
This   large  metallicity spread rules out a scenario in which the nuclear star cluster formed in a single burst of star formation.
The fraction of low-metallicity stars with $\mh$\,$\leq$\,$-$0.5\,dex  is (5.2$^{+6.0}_{-3.1}$)\,per cent.  
The spatial distributions and kinematics of metal-poor stars and   metal-rich stars are  similar. 
The metal-poor stars might originate from infalling globular clusters. However, our data set is dominated by metal-rich stars, and these stars have higher metallicities than the most metal-rich Milky Way globular clusters. A scenario in which  the nuclear star cluster is entirely formed from infalling globular clusters can be ruled out. 

 \label{sec:sec6}

 \section*{Acknowledgements}
RS acknowledges  funding from the European Research Council under the European Union's Seventh Framework Programme (FP7/2007-2013) / ERC grant agreement no. 614922.
This publication makes use of data products from the 2MASS, which is a joint project of the University of Massachusetts and the Infrared Processing and Analysis Center/California Institute of Technology, funded by the National Aeronautics and Space Administration and the National Science Foundation. 
This research made use of the SIMBAD data base (operated at CDS, Strasbourg, France).  We  thank Nikolay Kacharov,  and Iskren Georgiev for  discussions and suggestions.   Thanks to  Ariane Lan\c{c}on for helpful discussions. 
We finally thank the anonymous referee for   comments and suggestions. Based on observations collected at the European Organisation for Astronomical Research in the Southern Hemisphere, Chile (60.A-9450(A)).




\bibliographystyle{mn2e_trunc8} 
\bibliography{bibs_lt}



\appendix

\section{The $\teff$ -- $EW_\mathrm{CO}$ relation for giants}
\label{sec:tcorel}
We calibrated the relation of the effective temperature $\teff$ with the CO equivalent width $EW_\mathrm{CO}$ as defined by \cite{frogel} using the stars of the spectral library. We considered 69 stars with luminosity classes II--IV.  $EW_\mathrm{CO}$ was computed with $R= 3\,000$  for stars from \cite{wallacemr}, and $R=4\,350$ for the other stars in the spectral library.

The effective temperature $\teff$ is given for the stars in the GNIRS library  in  \cite{winge}, for the other stars we complemented $\teff$ with entries from the stellar parameter catalogue PASTEL \citep{pastel}. For stars with more than one entry we used the mean  of the various measurements, and the standard deviation as uncertainty $\sigma_{\teff}$. 
When there was no entry of $\teff$ for a star in the PASTEL catalogue, we used our knowledge of the spectral type. We used the list given in \cite{spectype}, where $\teff$ is listed  for various spectral types, and for dwarfs, giants, and supergiants separately. When necessary, we interpolated linearly between the spectral classes. We compared the   \cite{spectype} values for $\teff$ with the values given in tables 7.5, 7.6 and 7.7 in \cite{allen} and in table 15.7 in \cite{allen}. We used the scatter between these three different spectral type-$\teff$ tables to estimate the uncertainty $\sigma_{\teff}$.

\begin{figure}
  \centering  
  \includegraphics[width=0.99\columnwidth]{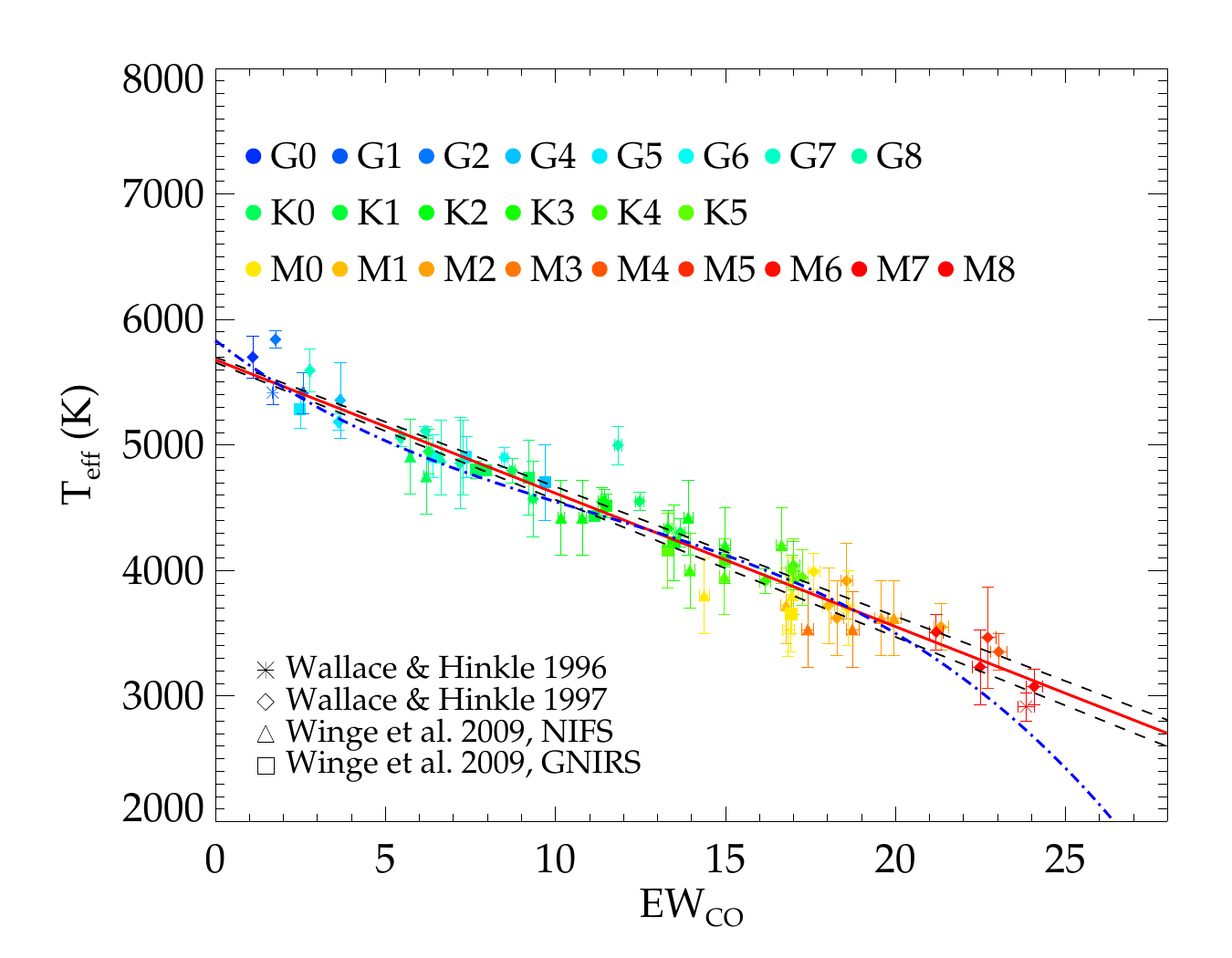}
 \caption{Relation between effective temperature $\teff$ and the equivalent width $EW_\mathrm{CO}$  for 69 template  stars with luminosity class II--IV.  
 The different colours denote different spectral types, the symbols denote the library from which the spectrum is taken.  The red line is a linear fit to the data, the black dashed lines show the formal uncertainty of the linear fit. The blue dot--dashed line is the $\teff$--$EW_\mathrm{CO}$ relation  as derived by \citet[]{pfuhl11}. }
 	\label{fig:cofrogel}
\end{figure}

Fig. \ref{fig:cofrogel} shows the relation between $\teff$ and $EW_\mathrm{CO}$ for 69 stars with luminosity classes II--IV from the spectral libraries. 
We made a linear fit to the data in the range below 6\,000\,K and found the relation
\begin{equation}
\teff = 5\,677 ^{\pm21}  \mathrm{K} - 106.3 ^{\pm 3.0}  \mathrm{K} \AA^{-1} \times EW_\text{CO},
\label{eq:teffco}
\end{equation}
where $EW_\mathrm{CO}$ is in $\AA$, and $\teff$ in K. The uncertainties are the formal fit uncertainties. The residual scatter is 163\,K. 
Our best-fitting result is shown as red line in Fig. \ref{fig:cofrogel}. We also show the three-degree-polynomial fit of \cite{pfuhl11} as blue dot--dashed line. \cite{pfuhl11}  used 33 giant stars from spectral libraries at lower spectral resolution ($R$\,$ \approx$\,3\,000 and $\approx$\,2\,000). Our $\teff$\,--\,$EW_\mathrm{CO}$ relation  is in agreement with their relation for  $\teff$\,$\ga$\,3\,000\,K.  However, at $\teff$\,$\la$\,3\,000\,K, the \cite{pfuhl11} relation declines faster than our linear fit. The different result of the relation can be explained by the larger sample and the   different values of $\teff$ that we use. We included  larger uncertainties $\sigma_{\teff}$ than \cite{pfuhl11}, since we took several measurements of $\teff$ for a star into account. We did not make a metallicity cut for the stars, since for many stars in the stellar library the metallicity is not known. When we just use a subsample of 39 spectra with near solar-metallicity in the range $-$0.5\,dex\,\textless\,$\mh$\,\textless\,+0.5\,dex, the relation changes within the formal fit uncertainties. 
\cite{2016A&A...590A...6S} showed that the $\teff$--$EW_\mathrm{CO}$  relation does not depend on the metallicity in the temperature range  3\,200--4\,500\,K and metallicities [$Fe/H$] from $-$1.2 to +0.5\,dex. We conclude that our $\teff$--$EW_\mathrm{CO}$  relation  is robust in this temperature and metallicity range.


\section{Table of stellar parameters}
\label{sec:tabsec}
\onecolumn
\begin{table}
\caption{Stellar parameters: stellar identification number Id, the coordinates in RA and Dec., the stellar parameters $\teff$, $\mh$, $\logg$, $v_z$, and extinction corrected $K_S$-band magnitude $K_{S,0}$.  The full table is available online.}
\label{tab:parameter}
\begin{tabular}{cccccccccccc}
Id & RA& Dec. & $\teff$ & $\sigma_{\teff}$ & $\mh$&$\sigma_{\mh}$ & $\logg$&$\sigma_{\logg}$ & $v_z$& $K_{S,0}$\\
&$ \left( ^{\circ} \right)$&$ \left( ^{\circ} \right)$&$ \left( \mathrm{K} \right)$&$ \left( \mathrm{K} \right)$&$ \left( \mathrm{dex} \right)$&$ \left( \mathrm{dex} \right)$&$ \left( \mathrm{dex} \right)$&$ \left( \mathrm{dex} \right)$&$\left( \mathrm{km}\,\mathrm{s}^{-1} \right)$&$\left( \mathrm{mag} \right)$&\\
\hline
$     1$&$  266.41675 $&$   -29.010296 $&$  3190 $&$  209 $&$   0.87 $&$   0.31 $&$    1.2 $&$    1.1 $&$     25.6 $&$   7.44 $ \\
$     5$&$  266.41571 $&$   -29.012167 $&$  3301 $&$  206 $&$   0.13 $&$   0.25 $&$    0.0 $&$    1.0 $&$   -241.4 $&$   7.88 $ \\
$     6$&$  266.42401 $&$   -29.003611 $&$  3374 $&$  205 $&$   0.14 $&$   0.25 $&$    0.0 $&$    1.0 $&$    133.5 $&$   7.79 $ \\
$    14$&$  266.42093 $&$   -29.004204 $&$  3250 $&$  217 $&$   0.31 $&$   0.24 $&$    0.1 $&$    1.0 $&$     69.5 $&$   7.78 $ \\
$    16$&$  266.41727 $&$   -29.013838 $&$  3200 $&$  205 $&$   0.31 $&$   0.25 $&$    0.0 $&$    1.0 $&$    -65.2 $&$   7.71 $ \\
$    17$&$  266.42236 $&$   -29.006290 $&$  3463 $&$  282 $&$  -0.37 $&$   0.24 $&$    0.2 $&$    1.0 $&$    115.2 $&$   7.87 $ \\
\hline
\end{tabular}
\end{table}

\bsp	
\label{lastpage}
\end{document}